\documentclass[manuscript]{acmart}
\usepackage{multirow}
\usepackage{xcolor}
\usepackage[table]{xcolor}
\usepackage{caption}
\usepackage{makecell}
\usepackage{booktabs}
\definecolor{uiblue}{HTML}{3B89F0}
\definecolor{uigreen}{HTML}{5EB761}
\definecolor{LA}{HTML}{DF8430} 
\definecolor{HA}{HTML}{4F88E6} 
\definecolor{LE}{HTML}{A83299} 
\definecolor{HE}{HTML}{50B948} 
\definecolor{UH}{HTML}{0B3D91}   
\definecolor{ENV}{HTML}{1B9E77}  
\usepackage{enumitem}

\AtBeginDocument{%
  }

\setcopyright{acmlicensed}
\copyrightyear{2018}
\acmYear{2018}
\acmDOI{XXXXXXX.XXXXXXX}
\acmConference[Conference acronym 'XX]{Make sure to enter the correct
  conference title from your rights confirmation email}{June 03--05,
  2018}{Woodstock, NY}

\acmISBN{978-1-4503-XXXX-X/2018/06}

\usepackage{multirow}
\usepackage{framed}
\usepackage{xcolor}
\usepackage[normalem]{ulem}
\usepackage{hyperref}
\usepackage{booktabs}

\begin{document}

\title{Tell Me Why, When, and How: Effects of Personality, Evidence, and Context on Older Adults’ Perceptions of Conversational AI Explanations}

\author{Niharika Mathur}
\email{nmathur9@jh.edu}
\authornote{Both authors contributed equally to the research}
\orcid{0000-0002-3969-7787}
\affiliation{%
  \institution{Johns Hopkins University}
  \city{Baltimore}
  \state{Maryland}
  \country{USA}
}

\author{Hasibur Rahman}
\email{rahman.has@northeastern.edu}
\authornotemark[1]
\orcid{0009-0008-0938-1504}
\affiliation{%
  \institution{Northeastern University}
  \city{Boston}
  \state{Massachusetts}
  \country{USA}
}

\author{Smit Desai}
\authornote{Corresponding author}
\email{sm.desai@northeastern.edu}
\orcid{0000-0001-6983-8838}
\affiliation{%
  \institution{Northeastern University}
  \city{Boston}
  \state{Massachusetts}
  \country{USA}
}

\renewcommand{\shortauthors}{Mathur et al.}
\renewcommand{\shorttitle}{Effects of Personality, Evidence, and Context}

\begin{abstract}
Large Language Model-based voice assistants (LLM-VAs) have shown potential to support older adults aging in place through proactive reminders, health information, and everyday assistance. As LLM-VAs become more conversational, designing explanations of their behavior requires understanding what information they communicate and how they deliver it. While prior work has studied older adults’ explainability requirements, relatively little is known about their perceptions of explanations when an LLM-VA’s conversational personality varies. To address this gap, we conducted a mixed-design study with 140 older adults, examining two dimensions of explanation design: evidential grounding (source of information) and the VA’s conversational personality (agreeableness and extraversion), across two contexts of use. Our results show that both dimensions influenced older adults' perceptions in distinct but interdependent ways, challenging one-size-fits-all explanation strategies. We conclude with practical design implications for coordinating explanation content, conversational delivery, and context in future LLM-VAs supporting aging in place.
\end{abstract}


  

\begin{teaserfigure}
  \includegraphics[scale=0.51]{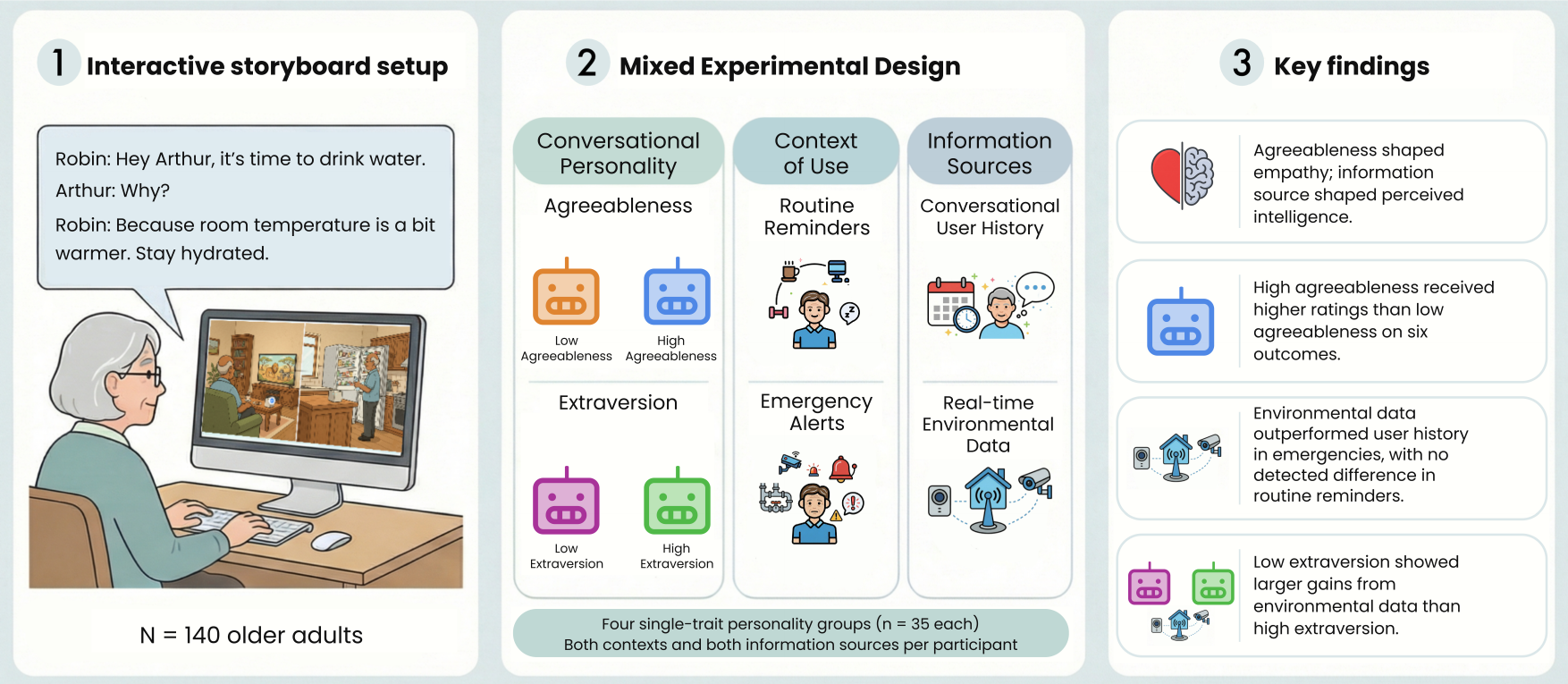}
  \centering
  \caption{Overview of our study of personality and explanations in LLM-based voice assistants (LLM-VA) for older adults ($N=140$). (1) Interactive storyboard setup: participants observed an at-home interaction in which an older adult (Arthur) interacted with an LLM-VA (``Robin'') delivering routine reminders or emergency alerts and explaining its behavior. (2) Experimental design factors. (3) Key findings.}
  \Description{Three panels read from left to right: study setup, experimental design, and key findings. The first depicts an older adult at a computer viewing an illustrated interaction between Arthur and the voice assistant Robin. Robin reminds Arthur to drink water; when Arthur asks why, Robin explains that the room is warm. The panel identifies 140 older adult participants. The middle panel separates conversational personality, context of use, and information source. Participants experienced one of four personality profiles—low or high agreeableness, or low or high extraversion—and both routine reminders and emergency alerts, with explanations based on either conversational user history or real-time environmental data. The final panel summarizes four findings: agreeableness affected perceived empathy while information source affected perceived intelligence; high agreeableness received higher ratings than low agreeableness on six outcomes; environmental data outperformed user history in emergencies, with no detected difference in routine reminders; and low extraversion showed larger gains from environmental data than high extraversion.}
\label{fig:teaser}
\end{teaserfigure}

\maketitle

\section{Introduction}

Imagine an older adult at home when their voice assistant (VA) suddenly reminds them to take their medication, warns them that the stove may have been left on, or alerts them to a possible emergency. Prior research in aging in place\footnote{The U.S. National Institutes on Aging (NIA) defines Aging in Place as ``home-based care that includes health, personal and other support services to help older adults stay in home environments as they grow older'' \cite{rogers2020defining}.} suggests that older adults could question the AI with responses such as: \textit{Why are you telling me this now?} \cite{enam2025artificial, gleaton2023understanding, mathur2025feels, zubatiy2023don} The assistant might answer using something the user told it previously or information gathered through sensors in the home, but what makes for an appropriate explanation may also depend on whether it is explaining a routine reminder or an immediate danger. Because that explanation is delivered conversationally, the assistant must also decide how to say it. A terse, reassuring, or enthusiastic explanation may cite the same evidence, but may not be received in quite the same way \cite{kim_should_2025}.

These questions become particularly important as VAs are increasingly explored as technologies for supporting older adults aging in place \cite{pradhan2020use, zubatiy2023don}. VAs can support reminders, health information, emergency alerts, and companionship \cite{desai2023painless, zubatiy2021empowering, mathur2022collaborative}. The integration of Large Language Models (LLMs) expands these possibilities further, giving rise to LLM-based voice assistants (LLM-VAs), which can sustain more natural conversations, draw on prior interactions and contextual information, and express different conversational personalities \cite{shahid_exploring_2025, kaliappan2024exploring}. These capabilities create considerably more flexibility in how explanations can be generated and delivered. An explanation grounded in a remembered preference may make sense for a routine reminder but be less convincing during an emergency, while the same conversational style may also be received differently depending on what the system is explaining and when. For older adults who may rely on these systems as part of everyday life, such differences can influence trust, social presence, continued use, critical evaluation of AI-generated information, and their sense of autonomy and agency \cite{barros2021circumspect, gallistl2024ai, mathur2026sometimes, chin2021being}.

The conversational richness of LLMs further exacerbates this challenge. LLM-VAs, such as Alexa, Gemini and ChatGPT Voice, are increasingly capable of sustaining multi-turn dialogue, referencing prior conversations and dynamically adapting responses to situational context \cite{mahmood2025user}, while also expressing specific personality traits \cite{rahman_vibe_2025}. A consequence of these advancements in LLM-VAs' flexibility and richness is that LLM-VAs are increasingly designed (and expected) to exhibit conversational personality traits \cite{sonlu2025effects, shahid_exploring_2025}. VAs are the voice-based subset of conversational agents (CAs), an umbrella term for text- and voice-based systems.\footnote{A conversational agent is a system designed to interact with humans using natural language, through text and/or spoken voice \cite{laranjo_conversational_2018, wei2023bot}.} In HCI, the Big Five framework \cite{john_big_1991, john_big-five_1999}, which characterizes personality along the dimensions of openness, conscientiousness, extraversion, agreeableness, and neuroticism, has been widely used to design conversational personalities for VAs \cite{volkel_user_2022}. Among these traits, \textit{agreeableness} and \textit{extraversion} have emerged as particularly perceptible and consequential, shaping perceptions of warmth, likeability, and social presence in CAs \cite{volkel_user_2022, volkel_examining_2021, ruane_user_2020, amin_kuhail_assessing_2024, papangelis2022understanding, rahman_vibe_2025}. These two traits are especially pertinent for older adults, who often evaluate VAs in interpersonal terms \cite{chin2024like, desai2023ok, chin2021being} and, at different times, view them as companions or information tools \cite{pradhan_phantom_2019}. Given this focus on the Big Five framework, we use the term \textit{conversational personality} to mean the deliberate expression of Big Five traits through pre-authored dialogue or natural-language prompting, as demonstrated in \cite{volkel_user_2022, volkel_examining_2021, ruane_user_2020, nass_does_2001, ramirez_controlling_2023, jiang_personallm_2024, serapio-garcia_personality_2025, rahman_vibe_2025}.

Prior HCI research has examined how conversational personality shapes user perceptions of CAs \cite{volkel_examining_2021, rahman_vibe_2025, kovacevic_chatbots_2024, shahid_exploring_2025, volkel_user_2022, ruane_user_2020, amin_kuhail_assessing_2024, papangelis2022understanding}. In parallel, recent work in Human-Centered Explainable AI (HCXAI) has examined how explanations can support user understanding, trust calibration and sensemaking, including how different sources of information can be leveraged to generate explanations \cite{das2023explainable, mathur_categorizing_2024}, and how explanation delivery is shaped by conversational properties such as AI persona, tone or timing \cite{mathur2026sometimes, he2025conversational}. However, these lines of work have largely treated explanation content and conversational personality separately. As a result, relatively little is known about how they might intersect, i.e., how the source of information grounding an explanation and the conversational personality of the LLM-VA delivering the explanation \textbf{together} shape older adults’ perceptions of the LLM-VAs for everyday assistive tasks in the home.

To address this gap, we examine these properties as two distinct \textit{layers} of explanation design, situated within a \textit{context of use}: (1) the information source grounding an explanation (the \textit{``evidential layer''}), capturing \textbf{why} the LLM-VA says it acted; (2) the conversational personality, operationalized using the Personality Trait Modulation Key Framework \cite{rahman_vibe_2025}, through which the explanation is delivered (the \textit{``conversational layer''}), capturing \textbf{how} it explains that action grounded in the \textit{context} in which the explanation is delivered, capturing \textbf{when} these explanations occur. 

To study these dimensions systematically, we conducted a mixed-design controlled study ($N=140$) in which older adult participants evaluated LLM-VA explanations through interactive storyboards (Fig. \ref{fig:teaser}). We manipulated the conversational layer \textbf{between participants} through four single-trait personality conditions: high or low agreeableness and high or low extraversion. For the evidential layer, we compared explanations grounded in two different information sources \textbf{within participants}: users' prior conversational history (UH) and real-time environmental data (ENV), adapted from the categorization introduced by \citet{mathur2026sometimes}. We additionally varied the context of use \textbf{within participants} between routine reminders and emergency alerts.

Our findings show that the evidential and conversational layers of LLM-VA explanations shaped older adults' evaluations in distinct but interdependent ways. High-agreeableness delivery was evaluated more favorably across information sources, while real-time environmental explanations were especially beneficial in emergency contexts and produced larger gains under low-extraversion delivery. Our results also highlight limitations in using rigid, \textit{one-size-fits-all} approaches to LLM-VA explanation design and suggest that effective explanations for older adults require coordinating what evidence a VA provides, how it communicates that evidence and the context in which the explanation is delivered.

We make three primary contributions to the design of LLM-VA explanations for older adults:

\begin{itemize}
    \item Conceptually, we demonstrate the potential for incorporating two complementary design dimensions for the design of LLM-VA explanations for older adults: the evidential grounding of an explanation and the conversational personality through which the explanation is delivered.
    \item Empirically, we provide evidence from a mixed-design study with older adults (\(N=140\)) showing that evidential grounding and conversational personality shape user perceptions in distinct but interdependent ways, with these effects varying across contexts of use.
    \item Practically, through these results, we offer tangible design implications for designers of LLM-VA explanations for older adults, including balancing the evidential content of an explanation, conversational personality and context of use in future LLM-VAs for aging in place.
\end{itemize}

\section{Related Work and Background}

\subsection{Voice Assistants for Older Adults Aging in Place}

A growing body of HCI research has demonstrated the potential of VAs to support older adults across a wide range of tasks, including medication management \cite{mathur2022collaborative, lee2015sensor}, cooking assistance \cite{chan2025mango}, household tasks \cite{chan2025insights}, informal learning \cite{desai2023ok, Desai_Chin_2021} and in providing companionship and support \cite{pradhan_phantom_2019, zubatiy2021empowering}. Beyond traditional smart speakers, voice-enabled interactions through humanoid robots \cite{chang2024dynamic, pino2015we}, interactive displays \cite{deusdad2025service} and ambient interfaces \cite{mynatt2001digital, consolvo2006design} have also been explored. Within the domain of assistive technologies for older adults, advances in ubiquitous computing and more recently, LLMs, have equipped VAs to leverage longitudinal conversational history and contextual information through integration with environmental sensors, wearable devices and smart home technologies, to provide more personalized and context-aware support to older adults \cite{lee2015sensor, el2020multimodal}, especially for safety-critical tasks such as fall detection and emergency response \cite{ejupi2016kinect, mohan2024artificial}. 

However, this conversational fluency and expanded contextual awareness also introduce parallel challenges for older adults’ understanding of these systems. As VAs become more proactive, personalized and capable of accessing diverse information sources, older adults must increasingly interpret why an AI assistant makes a recommendation or delivers an alert at a given moment \cite{zhang2025falling, mathur2023did}. \citet{gleaton2023understanding} found that older adults frequently questioned AI systems and expected usable explanations of AI behavior for sustained AI use. Additionally, concerns regarding the “black-box” nature of AI systems have also been shown to influence older adults’ willingness to engage \cite{enam2025artificial}. While research on AI explanations has shown their important role in influencing user trust and adoption \cite{scharowski2023exploring, pareek2024effect}, much of explainable AI (XAI) has traditionally focused on explaining complex model decisions to technical audiences \cite{ehsan2020human, miller2019explanation}, rather than supporting everyday users’ understanding within ongoing conversational interactions. This gap motivates much of our work, and we expand on the need for a human-centered perspective on AI explanations in the next subsection. 

\subsection{Evidential Grounding in Conversational AI Explanations}

Explainable AI (XAI) has traditionally focused on improving the transparency of AI models by exposing various aspects of a model’s internal reasoning \cite{arrieta2020explainable}, largely through techniques such as feature attribution, saliency maps, counterfactuals or local explanation frameworks (e.g., LIME, SHAP) \cite{dwivedi2023explainable, das2020opportunities}. While this has substantially advanced algorithmic transparency, the increasing integration of AI in everyday technologies in recent years has broadened the role of explanations beyond exposing model internals. As AI systems become embedded in healthcare, education and in homes, they increasingly support everyday, non-expert users who seek explanations to understand AI systems in a given context and make informed decisions \cite{wang2021explanations, kim2023help}. This focus on the human factors of explainability has motivated the emergence of Human-Centered Explainable AI (HCXAI) \cite{ehsan2020human, ehsan2023human}, which re-conceptualizes explanations as tools for supporting individual understanding, sensemaking, trust calibration and decision-making, rather than simply providing \textit{one-size-fits-all} technical explanations of computational processes. 

The emergence of LLMs further extends this human-centered perspective by embedding explanations within ongoing interactions between users and AI systems. As LLM-based systems increasingly interact through natural language, explanations become situated conversational interactions that unfold as part of ongoing dialogue \cite{mindlin2025beyond, bello2025three, mathur2026explanations}. Within this dialogue, explanations delivered as conversation can help users understand an AI’s recommendations and actions in context, while also allowing opportunities for clarification, negotiation, and adaptation through conversation. Early work on automated rationale generation demonstrated the benefits of natural language explanations in helping users form a stable mental model of AI behavior \cite{ehsan2019automated}, and more recent work has engaged with conversational explanations within dashboard-based interfaces. For example, \citet{he2025conversational} show how conversational XAI interfaces can support more natural and improved understanding, while also observing an overreliance on such interfaces, due to an “illusion of explanatory depth” conveyed through the AI's conversational personality. \citet{martens2025tell} demonstrate the importance of storytelling to provide narratives about AI predictions and report positive outcomes for conversational explanations. Regarding explanatory requirements, prior work has observed a difference in explanatory needs across user groups \cite{ehsan2023human, liao2021question, ridley2025human, xu2023hcai}, with developers seeking more technical explanations, while non-technical users often preferring conversational explanations embedded in their interaction goals \cite{huang2024research, shandilya2022understanding}. 

This body of work demonstrates a shift toward studying how explanations should be designed and communicated within conversation. While this presents an important motivating step, relatively little is known about the sources of information that conversational explanations should draw upon to help users understand the AI---a gap we aim to address in our work. This question is particularly important for LLM-VAs supporting older adults, a user group whose explanatory needs are grounded in contextual information present in their environment \cite{van2026should, gleaton2026taxonomy, mathur2026sometimes}, which we discuss next.

\subsubsection{Designing AI Explanations for Aging in Place}

Aging in place presents a particularly rich context for designing AI explanations, where interactions with VAs are shaped by the broader cognitive, social and environmental context in which those interactions occur \cite{kaliappan2024exploring}. Within the home, older adults often rely on a distributed network of physical and digital artifacts, such as calendars, smart displays etc., to support everyday activities and mediate technological interactions \cite{pradhan2023towards, pradhan2020use}. They also frequently draw upon these surrounding contextual resources to interpret and respond to a VA's reminders and recommendations \cite{mathur2025feels}, highlighting the need for explanations to account for this broader sociotechnical context.

Building on this, recent work has explored how information distributed within sociotechnical environments can be leveraged to generate more meaningful AI explanations. For example, conversational history and prior user interactions incorporated into explanations have shown to improve their usefulness and create a more holistic understanding of users beyond generic explanations \cite{baruah2024brief, alizadeh_voice_2024, mathur2026wants}. Similarly, frameworks that ground explanations in user activities (e.g., cooking, exercising, etc.) and their recurring activity patterns have also been proposed, enabling AI systems to justify recommendations using information directly relevant to users \cite{kofod2007explanations}. Smart homes and LLM-enabled environments now increasingly enable access to diverse forms of contextual information through stored conversational history, smart home sensors, wearable devices and other ambient technologies, creating new opportunities for designing AI explanations grounded in users' lived experiences and context \cite{das2023explainable, byrne2023good}. 

Particularly relevant to our goals in this paper is an exploratory study by \citet{mathur2026sometimes}, which examined how older adults perceived conversational explanations grounded in different forms of contextual evidence within smart home environments. In this study, older adults responded to the underlying source of information from which the AI appeared to derive its knowledge, using these "evidential cues" as explanations to reason its behavior. Inspired by their approach, we adopt a subset of these contextual evidence sources to structure the explanations in our study. Specifically, we examine conversational explanations grounded in either a user's \textbf{prior conversational history (UH)} or \textbf{real-time environmental information (ENV)}. Throughout this paper, we refer to this dimension as the \textit{evidential layer}, capturing \textit{what evidence} an explanation references when justifying a VA’s recommendation. 

While the evidential grounding of an explanation shapes \textit{why} the system says it acted, through the evidence it cites, conversational systems must also determine \textit{how} that explanation is communicated \cite{alves2026exploring, kopecka2024preferences}. Importantly, both may be received differently depending on \textit{when}, or in what context, the explanation is delivered, as routine and urgent situations can create different explanatory needs. The tone, language and agent personality accompanying an explanation can shape older adults' perceptions, particularly in long-term interactions. For example, \citet{okoso2025expressions} found that explanation tone (formal vs. humorous) influenced decision-making in scenarios where users turned to the AI for a “second-opinion”, while \citet{gomez2024large} found that conversational profile (rational vs. empathetic) had insignificant influence on the explanation acceptance. In our work, we conceptualize this as a complementary \textit{conversational layer} of explanation design, and review prior work on conversational personality next.

\subsection{Operationalizing the Conversational Layer Through Personality}
\label{sec:personality_va}

\textit{Conversational personality} is the perceived personality expressed by CAs through language \cite{rahman_vibe_2025}. For VAs, it is primarily expressed in spoken language. It includes the personality that users infer during interaction and the personality that designers deliberately shape through linguistic cues. Users form personality impressions of VAs even when personality is not experimentally assigned \cite{volkel_developing_2020, Kuzminykh_Sun_Govindaraju_Avery_Lank_2020}. Users also frame VAs in terms of social roles and metaphors \cite{desai_examining_2024,khadpe_conceptual_2020, pradhan_phantom_2019, desai2023metaphors}, while the CASA paradigm posits that people may apply social rules to computers without believing that they possess human feelings or intentions \cite{nass_computers_1994, gambino_building_2020}.

\subsubsection{Pre-LLM Conversational Personality Research}
Before LLMs, conversational personality research drew on the Big Five framework, expressing traits through speech, verbosity, word choice, self-disclosure, politeness, and directness \cite{john_big_1991, nass_does_2001, volkel_user_2022, ruane_user_2020}. Users distinguished personality profiles, and extraverted styles increased communication satisfaction, while agreeable agents were valued especially by agreeable users and disagreeable agents were broadly penalized \cite{volkel_user_2022, ahmad_extrabot_2020, volkel_examining_2021}. Similar patterns extended to VAs: users spontaneously described assistants in personality terms, and extraverted users preferred extraverted VAs, although introverted users showed no matching preference \cite{volkel_developing_2020, Snyder2023OneVoiceFitsAll}. These studies established personality as a consequential design choice, but fixed scripts limited expression to pre-authored profiles and manipulations of one or two traits \cite{ruane_user_2020, zhou_trusting_2019, volkel_examining_2021}.

\subsubsection{LLMs and Prompted Conversational Personality}

LLMs now allow shaping conversational personality through prompts rather than pre-authored responses. Researchers have used trait labels or trait-specific adjectives to prompt Big Five profiles and tested whether the LLM generated language expressed the intended traits \cite{ramirez_controlling_2023, jiang_personallm_2024, serapio-garcia_personality_2025}. PersonaLLM, for example, assigned descriptors for the high or low pole of all five traits \cite{jiang_personallm_2024}. Its authors then used Linguistic Inquiry and Word Count (LIWC) categories to assess whether the generated text exhibited corresponding psycholinguistic patterns \cite{jiang_personallm_2024, pennebaker_linguistic_2001}. Research also shows that users can perceive prompted Big Five profiles \cite{jiang_personallm_2024, kovacevic_chatbots_2024}, including when conversational personality is combined with embodiment \cite{sonlu2025effects}. Building on this work, \citet{rahman_vibe_2025} introduced Trait Modulation Keys (TMK) at CHI 2026, a modular prompting framework that pairs psychometrically grounded Personality Keys with Style Cues to manipulate conversational personality. We adopted TMK to design the LLM-VA personality; \S \ref{sec:design_personality} describes its implementation.

As commercial VAs integrate LLM backends \cite{mahmood2025user}, designers can increasingly shape how these systems communicate with users. Recent LLM-VA research has examined interaction breakdowns, healthcare communication, conversational style, and users' adaptation to generated speech \cite{mahmood2025user, yang_talk2care_2024, dittrich_conversational_style_2025, dittrich_mimicry_2026}. Research on LLM-VAs for older adults has also explored conversational personality preferences. Specifically, \citet{shahid_exploring_2025} elicited agreeableness, extraversion, and conscientiousness in an LLM-VA. Older adults distinguished the resulting personality profiles, although their preferences differed between brief laboratory interactions and longer-term home use. This work focused on companionship rather than explanations of system actions.

\subsubsection{Agreeableness and Extraversion}
Across text-based CAs and VAs, agreeableness and extraversion are among the most perceptible conversational personality traits \cite{volkel_user_2022, volkel_examining_2021, ruane_user_2020, amin_kuhail_assessing_2024, papangelis2022understanding, rahman_vibe_2025}. Agreeableness is characterized by cooperation, warmth, and social responsiveness \cite{john_big-five_1999, goldberg_development_1992, mccrae_introduction_1992}. Agreeable, caring, or polite communication can shape users' preferences, trust, and acceptance of CAs and VAs \cite{zhong_effects_2022, volkel_examining_2021, chin2024like, wong_voice_2024}. However, friendliness can undermine trust when it becomes sycophantic \cite{sun2025friendly}. Extraversion is characterized by talkativeness, assertiveness, and energetic engagement \cite{john_big-five_1999, goldberg_development_1992}. Extraverted CA styles can increase communication satisfaction \cite{ahmad_extrabot_2020}, but extraverted language is often more verbose \cite{pennebaker_linguistic_2001, gill2019taking}. Excessive verbosity in LLM output can introduce repetition and ambiguity and is associated with poorer task performance \cite{zhang_demystify_2025}.

These traits are pertinent for older adults, who often evaluate VAs in interpersonal terms \cite{chin2024like, chin2021being} and may view them as companions or information tools \cite{pradhan_phantom_2019}. In a longitudinal pilot, older adults' extraversion predicted routine initiation, whereas their agreeableness predicted perceived usability \cite{shade_evaluating_2025}. Prior work has focused on conversation, acceptance, or companionship. We therefore examine how an LLM-VA's agreeableness or extraversion shapes older adults' perceptions of its explanations and whether these effects vary with the information source supporting an explanation.

\subsection{Bringing Conversational and Evidential Layers Together}

HCXAI research has focused on the information that explanations should include, such as contextual grounding, user relevance, and alignment with everyday tasks \cite{ehsan2020human, ferreira2020people}. Research on VAs has separately examined how personality traits and interaction styles shape warmth, competence, trust, and system evaluation. LLM-VAs bring these two concerns together: they can draw on different sources of contextual information while also modulating how explanations are conversationally delivered \cite{bello2025three, alizadeh_voice_2024}. Human-Centered Conversational XAI further positions properties such as persona and voice as part of the explanation itself, suggesting that the same underlying information may be received differently depending on how it is communicated \cite{mathur2026explanations}.

Building on these perspectives, we conceptualize conversational AI explanations as comprising two complementary design dimensions: an \textit{evidential layer} (why), concerning what information grounds an explanation, and a \textit{conversational layer} (how), concerning how that explanation is communicated. Importantly, the effectiveness of these layers may also depend on the \textit{context} (when) in which an explanation is delivered. We therefore examine their independent and interactive effects by manipulating the LLM-VA's conversational personality, the information source grounding its explanations, and the context of use. We ask:

\begin{itemize}
    \item \textbf{RQ1:} How do the conversational layer of LLM-VA explanations, operationalized through \textit{agreeableness} and \textit{extraversion}, and the evidential layer, operationalized through \textit{Conversational User History} and \textit{Real-time Environmental Data}, independently shape older adults' perceptions of those explanations?

    \item \textbf{RQ2:} How do the conversational and evidential layers interact, and how do their effects vary across \textit{routine-reminder} and \textit{emergency-alert} contexts of use?
\end{itemize}

\noindent RQ1 examines the independent effects of the conversational and evidential layers, while RQ2 examines how these layers interact and how their effects vary across contexts of use.

\section{Method}
\label{sec:study_design}

We conducted an Institutional Review Board (IRB)-approved mixed-design experiment ($N=140$) to examine how older adults perceived explanations from an LLM-VA.

\subsection{Design Overview}

An explanation from an LLM-VA varies in both how the assistant speaks and in the evidence it cites. Recent work treats such conversational explanations as dialogues whose reception depends on properties such as persona, tone, timing, pacing, repair, and interaction history \cite{mathur2026explanations}. We separated the two functions of LLM-VA's explanation into a \textbf{conversational layer} and an \textbf{evidential layer}, and we evaluated both inside a \textbf{context of use} (See Figure \ref{fig:overview}). Each of the three elements became one factor in the design:

\begin{figure*}[ht]
    \centering
    \includegraphics[width=\textwidth]{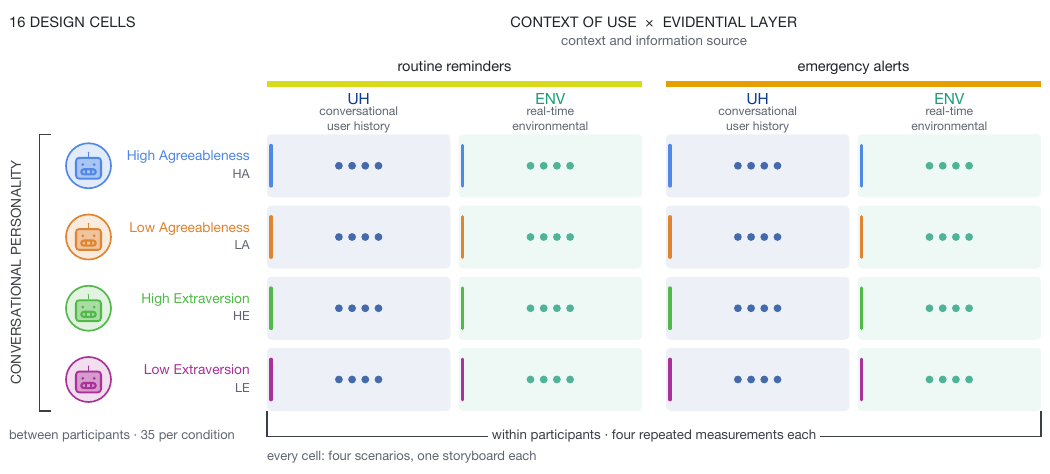}
    \caption{\textbf{Overview of the mixed experimental design}. The conversational personality of the LLM-based voice assistant (LLM-VA) varied between participants through four conditions: High Agreeableness (HA), Low Agreeableness (LA), High Extraversion (HE), and Low Extraversion (LE; $n=35$ each). Context of use (routine reminders or emergency alerts) and information source (Conversational User History [UH] or Real-time Environmental Data [ENV]) varied within participants. Each participant viewed four storyboards in each context $\times$ information source combination (16 total). Participants provided one set of ratings after each combination (four repeated measurements).}
    \Description{A four-row, four-column matrix represents 16 experimental cells. From top to bottom, the rows are high agreeableness, low agreeableness, high extraversion, and low extraversion. The columns are routine reminders using conversational user history, routine reminders using real-time environmental data, emergency alerts using user history, and emergency alerts using environmental data. Four dots inside each cell represent four storyboard scenarios. A bracket beside the rows identifies personality as a between-participant factor, with 35 participants assigned to each profile. A bracket beneath the columns identifies context and information source as within-participant factors: each participant experienced all four columns in their assigned personality row. Thus, each participant viewed 16 storyboards and completed four sets of ratings, one after each group of four scenarios.}
    \label{fig:overview}
\end{figure*}

\begin{itemize}[leftmargin=*]
 \item \textbf{The context of use} is the situation the explanation arrives in. We manipulated it \textit{within participants} through the urgency and risk of the assistant's action: a routine reminder or an emergency alert (\S \ref{sec:context}).
  \item \textbf{The evidential layer} is the information an explanation cites as evidence for the assistant's action. We manipulated it \textit{within participants} through the source of that information: Conversational User History (UH) or Real-time Environmental Data (ENV) (\S \ref{sec:info_source}).
    \item \textbf{The conversational layer} is how the assistant delivers an explanation. We manipulated it \textit{between participants} through the assistant's \textit{conversational personality}, which is Big Five traits deliberately expressed in its explanations. Each participant heard one of four single-trait profiles: High Agreeableness (HA), Low Agreeableness (LA), High Extraversion (HE), or Low Extraversion (LE) (\S \ref{sec:design_personality}).
\end{itemize}

Separating the two layers requires each factor to vary on its own, so we defined what each factor controlled. The information source determined which facts an explanation cited. The conversational personality condition determined how those facts were worded. Within a scenario and information source, all four personality conditions cited the same facts; within a scenario and personality condition, the UH and ENV explanations differed only in the evidence they cited (Table \ref{tab:condition_layout}). We held the rest of the conversational layer constant. Every explanation was a single turn delivered at the same point in a fixed interaction sequence, so timing and pacing did not vary (\S \ref{sec:storyboard}), and we synthesized all audio with one voice at a neutral emotion setting and the default pitch, so vocal affect did not vary (\S \ref{sec:voice}). Our conversational-layer claims, therefore, concern conversational personality rather than the layer as a whole.

The study used a mixed design with four personality conditions between-subjects (HA, LA, LE, and HE) and two factors crossed within-subjects: source of information and context. Each personality group experienced all four combinations of information source and context, yielding 16 condition combinations across the study and four repeated measurements per participant. We counterbalanced the order of the two contexts, the order of the two information sources within each context, and the order of scenarios within each combination. Personality condition entered the analysis as a four-level factor, and the HA vs.\ LA and HE vs.\ LE contrasts, which isolate agreeableness and extraversion respectively, provided the trait-level inference (\S \ref{sec:data_analysis}).

We delivered all conditions through narrative vignettes within interactive storyboards featuring a fictional LLM-VA named \textit{Robin}, presented as a box-like, in-home voice assistant that gave reminders and alerts for household tasks. Robin's utterances were presented as pre-recorded audio stimuli, a common approach in VA research that allowed us to keep the voice, pacing, wording, and timing of the conversational exchanges consistent across participants and conditions, avoiding variation introduced by factors such as latency, responsiveness, or speech generation \cite{desai_examining_2024, Desai_Chin_Wang_Cowan_Twidale_2026, kim_should_2025}. We supplemented these audio stimuli with interactive storyboards to make the scenarios more vivid and easier to understand, particularly for older adult participants, while retaining the experimental consistency afforded by controlled scenario-based evaluation \cite{chang2024dynamic, chang2025unremarkable, luria2020social}. The storyboards placed each exchange within a daily routine and showed what was happening when Robin intervened, giving participants additional context for interpreting the reminder or alert and the explanation that followed. The interface, narrative structure, and assistant's appearance remained constant across conditions, while the explanation characteristics required by the experimental design were varied systematically. Grounding these interactions in familiar routines and everyday narratives further helped participants imagine the experimental situations more clearly \cite{truong_storyboarding_2006, go2003scenario, dow_external_2006}.

We built these stimuli in three steps. We designed two contexts of use, four scenarios in each, and an explanation for every scenario under both information sources and all four personality conditions. The following subsections describe each step.

\subsection{Designing the Contexts of Use}
\label{sec:context}

The context of use had to vary in urgency and risk while remaining part of ordinary home life. Prior work on smart-home technologies and conversational AI identifies the everyday activities in which older adults use or envision AI support \cite{zubatiy2021empowering, pradhan_phantom_2019, mathur2022collaborative}. In interviews with older adults, \citet{chan2025insights} identified two occasions for AI intervention in aging in place: routine-based reminders, which concern recurring habits and preferences, and real-time safety triggers, which concern situations requiring immediate action. We adopted this pair as our two levels. \textbf{Routine reminders} were low-risk, recurring interactions concerning meals, exercise, or errands. \textbf{Emergency alerts} were high-risk, time-sensitive situations involving a potential hazard and immediate action, including fire hazards, health incidents, and security concerns. The two levels, therefore, differ in recurrence, urgency, and consequence, the properties we intended to vary.

We treated context as a factor. Older adults value different explanations in different circumstances, and the conversational appropriateness of an explanation shifts with the stakes \cite{mathur2026sometimes, mathur2026explanations}. A human-centered account of aging-in-place environments makes the same point: the relevance of a piece of information depends on the user's goals, routines, and task context \cite{mathur_categorizing_2024}.

\subsubsection{Designing the Scenarios}
\label{sec:scenarios}

A scenario in our study was a short household situation in which \textit{Robin} acted and then explained itself. \textit{Robin} gave Arthur, an older adult living alone, a reminder or an alert, and Arthur could ask why. Each scenario had to meet three requirements. It had to belong to one of the two contexts. It had to support a credible explanation from either information source, so that a user-history basis and a sensor basis were both plausible grounds for the same action. And it had to be a situation in which a user might reasonably ask why the assistant acted, because an explanation carries information only when the behavior is not self-evident \cite{wolf2019explainability}.

We generated candidates following Go and Carroll's guidelines for scenario-based ideation in HCI \cite{go2003scenario}, starting with 50 short prompts. Research on aging in place guided the activities, including Activities of Daily Living (ADLs) and Instrumental Activities of Daily Living (iADLs) \cite{katz1983assessing, gold2012examination, graf2008lawton, pfeffer1982measurement}. We also drew on field studies of smart-home voice-based AI interactions with older adults \cite{pradhan2020use, zubatiy2021empowering, mathur2026sometimes, chin2024like, desai2023ok}. This research identifies reminders for daily activities, appointments, groceries, and food as common uses, as well as alerts about physical dangers in the home \cite{zubatiy2021empowering, zubatiy2023don}.

We then applied the three requirements to the candidates and dropped those that repeated an activity already covered. Eight scenarios remained, four routine reminders and four emergency alerts. Four scenarios per context, rather than one, reduced the chance that a single situation drove the ratings \cite{schoemaker1993multiple}. We wrote each of the eight under both information sources, producing the 16 storyboards that every participant saw.

\subsection{Designing the Evidential Layer: Information Source}
\label{sec:info_source}

The evidential layer is the information an explanation offers as evidence for the assistant's action, so we operationalized it as the source of that information. \citet{mathur2026sometimes} categorize four types of in-home information that an AI explanation can draw on: conversational history, environmental data, activity-based inferences, and internal system logic. We built our two conditions from the first three. \textbf{Conversational User History (UH)} explanations cited prior conversations and stated user preferences, such as a routine the user had described. \textbf{Real-time Environmental Data (ENV)} explanations cited current sensor and wearable signals together with an activity-based inference that a Human Activity Recognition system drew from those signals. We combined environmental data with activity-based inferences in a single condition because the two co-occur in deployed systems, and older adults describe both as concrete and situational \cite{mathur2026sometimes}. We excluded internal system logic because our scenarios explained the contextual basis for an action.

Both conditions, therefore, name the contextual information behind a reminder or an alert instead of exposing model internals through counterfactual reasoning or feature attribution. They differ in where that information came from, when the system gathered it, and how it framed the user. Table \ref{tab:information_sources} summarizes the two operationalizations. Within a scenario, the two versions justified the same action with different evidence, and we kept the number of cited facts and the decision they supported comparable across sources.

\begin{table*}[ht]
\centering
\caption{The two information source conditions used in our study: Conversational User History (UH) and Real-time Environmental Data (ENV), adapted from \citet{mathur2026sometimes}.}
\label{tab:information_sources}
\begin{tabular}{p{0.20\linewidth}|p{0.36\linewidth}|p{0.36\linewidth}}
\toprule
 & {\centering\cellcolor{UH!15}\textbf{\textcolor{UH}{Conversational User History (UH)}}\par} & {\centering\cellcolor{ENV!15}\textbf{\textcolor{ENV}{Real-time Environmental (ENV)}}\par} \\
\hline
\textbf{Information source} &
Prior conversations and user-specific history (e.g., preferences, routines). &
Real-time sensor or wearable signals and activity-based inferences drawn from them. \\
\hline
\textbf{Temporal orientation} &
Past-oriented: references what the user previously said or did. &
Present-oriented: references what the system currently detects or infers. \\
\hline
\textbf{Framing style} &
Personalized and relational. &
Factual and situational. \\
\bottomrule
\end{tabular}
\end{table*}

\subsection{Designing the Conversational Layer: Prompting Personality with TMK}
\label{sec:design_personality}

The conversational layer is how the assistant delivers an explanation, so we operationalized it as the assistant's \textit{conversational personality}, Big Five traits deliberately expressed in the language of its explanations. 

\subsubsection{Trait Modulation Keys (TMK)}
To manipulate conversational personality, we adopted TMK, a modular prompting framework introduced by \citet{rahman_vibe_2025} at CHI 2026. TMK pairs two complementary sets of instructions for each Big Five trait: \textit{Personality} instructions specify the intended trait expression, while \textit{Style Cues} guide how that expression is conveyed through language. These paired instructions are provided at three expression levels—high, medium, and low—allowing traits and their intensity to be configured separately. In our study, we used the low- and high-expression keys for the most readily perceived traits ( as detailed in \S \ref{sec:personality_va}): agreeableness and extraversion. The framework builds on evidence that trait descriptions can guide personality expression in LLM-generated responses~\cite{jiang_personallm_2024, serapio-garcia_personality_2025}, providing a validated basis for operationalizing our personality conditions.

A \textit{Personality Key} specifies the agent's intended disposition using trait descriptors drawn from psychometric resources. A \textit{Style Cues Key} translates that disposition into instructions for tone, word choice, sentence structure, and conversational markers, informed by research on personality expression in language \cite{rahman_vibe_2025}. For example, our HA Personality Key described Robin as ``trustful, kind, considerate, and warm,'' while the corresponding Style Cues Key instructed it to use politeness markers, soften suggestions, and use cooperative pronouns. The LA keys paired an uncooperative disposition with blunt directives and minimal hedging. The HE keys paired an energetic, talkative disposition with enthusiastic wording and invitations to interact; the LE keys paired a reserved disposition with concise, factual sentences and minimal small talk. Together, the keys specified both the intended trait and concrete ways to express it in an explanation.

Following TMK's prompt architecture, we combined the assigned keys with the task instructions and explicitly directed GPT-5 to follow the Personality Key for behavioral decisions and the Style Cues Key for communication style \cite{rahman_vibe_2025}. \citet{rahman_vibe_2025} validated TMK across all 243 Big Five trait-level configurations (low, medium, and high; $3^5$). The validation reported a 92.3\% targeted trait match rate, internal consistency of $\alpha \ge .973$, convergent and discriminant validity of $\Delta = .922$, and theoretically consistent relationships with external measures. We further validated TMK across our 4 conditions for GPT-5, achieving 100\% accuracy and convergent validity above $\Delta = 0.99$ (See Figure \ref{fig:tmk}). The supplementary materials provide the generation prompt and condition-specific keys. To assess whether participants perceived the intended traits in the generated explanations, we conducted a separate manipulation check; its procedure and results are reported in \S \ref{sec:manipulation_check}.

\subsubsection{Explanation Generation}

\begin{table*}[ht]
\centering
\caption{Content-matched examples illustrating the four LLM-VA personality conditions (High/Low Agreeableness, High/Low Extraversion) under the two information sources defined in Table \ref{tab:information_sources}. Within each information-source row, all four examples cite the same source-specific facts for the same routine trash-day reminder; only the conversational realization varies.}
\label{tab:condition_layout}
\begin{tabular}{p{0.14\linewidth}|p{0.39\linewidth}|p{0.39\linewidth}}
\toprule
& {\centering\cellcolor{HA!15}\textbf{\textcolor{HA}{High Agreeableness (HA)}}\par}
& {\centering\cellcolor{HE!15}\textbf{\textcolor{HE}{High Extraversion (HE)}}\par} \\
\hline
\textbf{Conversational User History (UH)}
& \textit{You told me that you prefer to take the trash out on Thursday mornings after breakfast. Since it is Thursday morning, I thought a reminder now might be helpful.}
& \textit{You told me you prefer to take the trash out on Thursday mornings after breakfast, and it is Thursday morning---perfect time to get it done!} \\
\hline
\textbf{Real-time\newline Environmental (ENV)}
& \textit{The kitchen motion sensor detected activity this morning, and the trash-bin sensor shows the bin is full. I thought a reminder while you are nearby might be helpful.}
& \textit{The kitchen motion sensor detected activity this morning, and the trash-bin sensor shows the bin is full---perfect timing to take it out while you are nearby!} \\
\midrule
& {\centering\cellcolor{LA!15}\textbf{\textcolor{LA}{Low Agreeableness (LA)}}\par}
& {\centering\cellcolor{LE!15}\textbf{\textcolor{LE}{Low Extraversion (LE)}}\par} \\
\hline
\textbf{Conversational User History (UH)}
& \textit{You said you prefer to take the trash out on Thursday mornings after breakfast. It is Thursday morning, so take it out now.}
& \textit{You prefer Thursday mornings after breakfast for taking out the trash. It is Thursday morning.} \\
\hline
\textbf{Real-time\newline Environmental (ENV)}
& \textit{The kitchen motion sensor detected activity this morning, and the trash-bin sensor shows the bin is full. Take it out now.}
& \textit{The kitchen motion sensor detected activity this morning. The trash-bin sensor shows a full bin.} \\
\bottomrule
\end{tabular}
\end{table*}

The TMK prompting framework produced the profiles in Table \ref{tab:condition_layout}, which holds the cited facts constant within each information source. The High Agreeableness Personality Key describes the agent as ``kind, considerate, and warm,'' while its Style Cues Key adds politeness markers and hedges, such as ``I thought a reminder now might be helpful.'' The Low Agreeableness keys produce directive phrasing: ``It is Thursday morning, so take it out now.'' The High Extraversion keys produce socially expressive language, such as ``perfect time to get it done!'' The Low Extraversion keys produce terse, matter-of-fact sentences: ``The trash-bin sensor shows a full bin.''  
Table \ref{tab:condition_layout} shows the four personality profiles under both information-source conditions from Table \ref{tab:information_sources} for one routine trash-day reminder.

We applied TMK's validated Personality Keys and Style Cues Keys to configure \textit{Robin}'s Agreeableness and Extraversion at high and low levels. We used GPT-5 to generate the explanations from the same scenario materials, the assigned interaction context, the information source (UH or ENV), and the keys for the target personality condition. Only the scenario context changed context-specific information, and only the information source changed source-specific information. The researchers reviewed the explanations for decision-relevant informational comparability, clarity, and appropriate trait expression.

\subsection{User Interface Development}
\subsubsection{Interactive Storyboard}
\label{sec:storyboard}
Storyboards can require users to infer unspecified parts of an interaction from a narrative artifact \cite{kim_sketchstudio_2018, benford_interaction_2009}. \citet{dow_external_2006} argue for interactive tools that can support or be embedded into storyboards. We therefore embedded audio clips in an interactive storyboard rather than using the audio-only stimuli of prior LLM-VA studies \cite{desai_examining_2024, pias_impact_2024, jamshed_designing_2025}. This format presented each explanation in context.

\begin{figure*}[ht]
    \centering
    \includegraphics[scale=0.66]{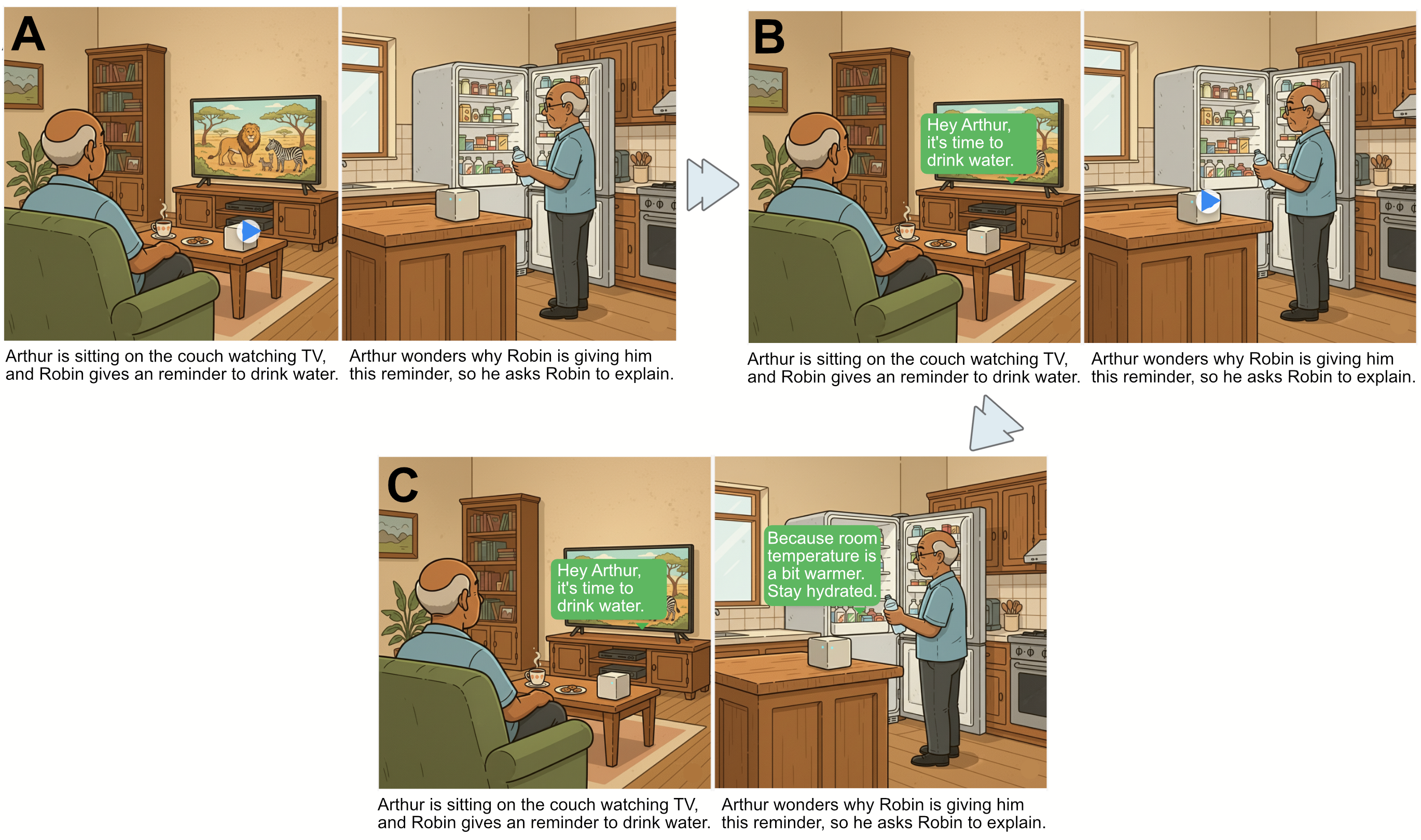}
    \caption{Interactive storyboard UI and sequence (A$\rightarrow$B$\rightarrow$C). Participants clicked the \textcolor{uiblue}{\Large$\blacktriangleright$} button to play the reminder or alert (A) and then the explanation (B). After each audio clip, a \textcolor{uigreen}{\Large$\blacksquare$} speech bubble displayed the transcript (B)(C).}
    \label{fig:xai}
    \Description{Three interface states, labeled A, B, and C, are connected by arrows in that order. Each state contains the same two illustrations: Arthur watching television on the left and standing beside an open refrigerator on the right, with Robin represented by a small speaker. In A, a triangular play button on the left starts Robin's reminder. In B, the reminder has finished and a speech bubble displays “Hey Arthur, it's time to drink water.” A second play button on the right starts Robin's explanation. In C, both transcripts remain visible; the right speech bubble reads “Because room temperature is a bit warmer. Stay hydrated.” The sequence shows how participants heard the reminder before its explanation, with each transcript appearing after its corresponding audio clip}
\end{figure*}

Our storyboards were implemented using a web-based, interactive storybook interface (Fig. \ref{fig:xai}) with two illustrated panels. The left panel established the context, such as Arthur sitting on the couch watching TV. The right panel depicted \textit{Robin} delivering its explanation. We generated the illustrations with generative AI in a consistent cartoon style to focus attention on the interaction rather than photorealistic detail \cite{truong_storyboarding_2006}.

Interaction followed a fixed A$\rightarrow$B$\rightarrow$C sequence. In step (A), participants clicked the \textcolor{uiblue}{\Large$\blacktriangleright$} button on the left panel to hear \textit{Robin}'s reminder or alert (e.g., \textit{
``Hey Arthur, it's time to drink water''}). When the audio ended, a \textcolor{uigreen}{\Large$\blacksquare$} speech bubble displayed the transcript in step (B). Participants then clicked a second \textcolor{uiblue}{\Large$\blacktriangleright$} button on the right panel to hear the explanation (e.g., \textit{``Because room temperature is a bit warmer. Stay hydrated''}). A \textcolor{uigreen}{\Large$\blacksquare$} speech bubble displayed this transcript in step (C). Participants could not skip ahead, and the audio with persistent transcripts supported different processing preferences and comprehension.

\subsubsection{Voice Design}
\label{sec:voice}
We generated all audio stimuli with Cartesia's Sonic-2\footnote{\url{https://cartesia.ai/sonic}} text-to-speech engine. We selected ``Jacqueline,'' a female American English voice. Older adults prefer female synthetic voices \cite{voicebotai_what_2019} and show greater speech alignment toward female human and device voices than younger adults \cite{zellou_age-_2021}. We used a neutral emotion setting and the default pitch in all conditions so that vocal affect did not confound the personality manipulation. Because older adults' speech comprehension declines as speech rate increases \cite{sutton_younger_1995, schneider_speech_2005, lee_aging_2015, huang_designing_2025}, we slowed the audio by 10\%. We avoided unnaturally slow speech, which can be perceived as patronizing \cite{kemper_experimentally_1999}.

The supplementary materials include all eight storyboard scenarios and the 72 audio files.

\subsection{User Perception Measures}
We used seven measures spanning two dimensions, based on a classification of CUI measures \cite{wei2023bot} and explanation-effectiveness measures from \citet{hoffman2023measures}. The first dimension concerned perceptions of \textit{Robin} while it explained reminders and alerts. The second concerned the explanations themselves.

\begin{enumerate}
    \item \textbf{Perception of LLM-VA as it delivered its explanations:} Prior work in human-AI interaction in domestic settings like the home shows the expectation to understand users' lived experiences, often attributed to a sense of perceived empathy \cite{forlizzi2007robotic, cagiltay2024methods, roshanaei2025talk}. We measured \textit{perceived empathy} with the 6-item Emotional Responsiveness factor of PETS \cite{schmidmaier2024perceived} (101-point slider). Given the importance of sustained VA adoption for older adults \cite{upadhyay2023studying}, we measured \textit{intention to adopt} using a 2-item scale adapted from \citet{moussawi2021perceptions} (7-point Likert scale). We measured \textit{trustworthiness} with the 5-item TiA questionnaire \cite{jian2000foundations} (5-point Likert) and \textit{perceived likeability} and \textit{intelligence} with two 5-item Godspeed subscales \cite{bartneck2023godspeed, bartneck2009measurement} (5-point semantic differential).
    \item \textbf{Perception of the explanations themselves:} Explanations shape trust, confidence, and reliance in human-AI interactions \cite{edmonds2019tale, bussone2015role} and can reduce overreliance \cite{lammert2024humans, scharowski2023exploring, vasconcelos2023explanations}. Because the storyboard design precluded behavioral observation, we adapted the 5-item reliance questionnaire from Lyons and Stokes \cite{lyons2012human} to measure \textit{user reliance}. We report the results of an exploratory factor analysis in the Results section. We measured \textit{explanation satisfaction} with \citet{hoffman2023measures}'s 7-item scale (5-point Likert), defined as the degree to which users feel they understand the AI system being explained. We used this measure to evaluate the explanations as conversational justifications of the LLM-VA’s behavior, used similarly in prior work \cite{xu2025confronting, pareek2025s, martin2026papertrail}.
\end{enumerate}

We also measured participants' agreeableness and extraversion with a 4-item domain-specific Mini-IPIP scale\footnote{The Mini-IPIP (Mini-International Personality Item Pool) measures the Big Five personality traits \cite{donnellan_mini-ipip_2006}.} (5-point Likert).

\subsection{Procedure}

\begin{figure*}[ht]
    \centering
    \includegraphics[width=\textwidth]{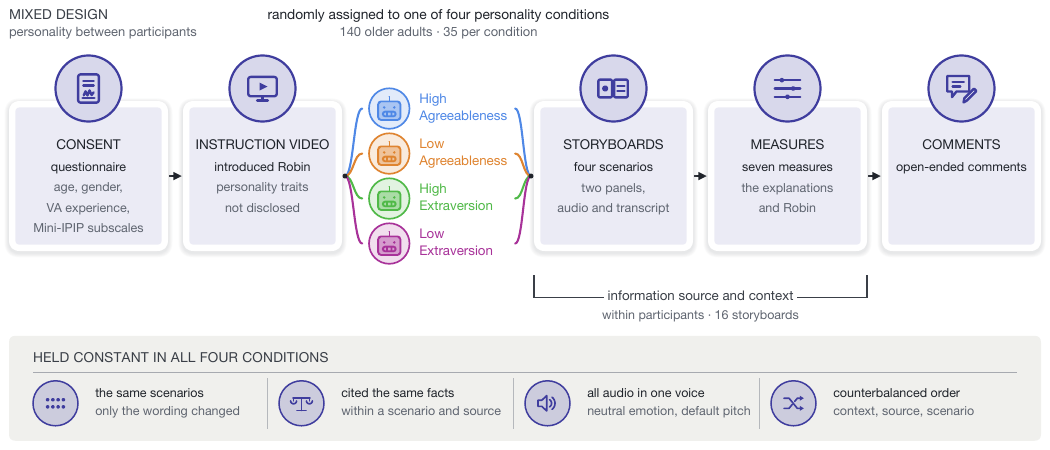}
    \caption{\textbf{Study procedure.} After consent and questionnaires on demographics, voice-assistant experience, and participant agreeableness and extraversion, an instruction video introduced \textit{Robin} without disclosing its personality. We then randomly assigned 140 older adults to one of four conversational-personality conditions ($n=35$ each). Participants viewed 16 interactive storyboards across context and information source, completed seven perception measures after each four-scenario block, and provided an optional open-ended comment in the end. Within each scenario and source, the four personality conditions cited the same facts; only the wording changed. We also used one synthesized voice with neutral emotion and default pitch and counterbalanced the presentation order of contexts, sources, and scenarios.}
    \label{fig:procedure}
    \Description{A flowchart runs from left to right through consent and questionnaires, an instruction video, random personality assignment, storyboards, outcome measures, and comments. The questionnaires cover age, gender, voice-assistant experience, and participant personality. The video introduces Robin without disclosing its personality. The flow then branches into high agreeableness, low agreeableness, high extraversion, and low extraversion, with 35 participants per branch, before rejoining at the storyboard stage. A bracket links storyboards and measures: participants view four scenarios and complete seven measures for each context and information-source combination, totaling 16 storyboards. Optional open-ended comments come last. A strip beneath the flow lists the design controls: the same scenarios, the same cited facts within each scenario and source, one synthesized voice with neutral emotion and default pitch, and counterbalanced presentation order. Personality conditions vary the wording of the explanations.}
\end{figure*}

We recruited participants through Prolific\footnote{\url{https://www.prolific.com/}}. After providing informed consent, they completed a demographic questionnaire (age, gender, VA experience) and the Mini-IPIP Agreeableness and Extraversion subscales. An instruction video introduced \textit{Robin} but did not disclose its personality traits, to avoid anchoring effects \cite{rhue2023anchoring, desai_examining_2024}. We randomly assigned each participant to one of four personality conditions: high or low agreeableness or high or low extraversion.

Participants viewed storyboards of \textit{Robin} assisting an older adult in routine reminders and emergency alerts (Fig. \ref{fig:xai}). Each context comprised four scenarios, and participants saw every scenario under both information sources. Each participant, therefore, viewed 16 storyboards: eight scenarios, each with a UH and an ENV explanation. Four scenarios per information source reduced bias from exposure to a single scenario \cite{schoemaker1993multiple}. After four scenarios of one information source, participants completed the seven measures for the explanations and \textit{Robin}. They then viewed the other information source in the same context and repeated the process in the second context. Each participant provided four sets of ratings, one for each context and information-source combination. We counterbalanced this sequence as described in \S \ref{sec:study_design}. Two attention checks assessed attention and response accuracy. After all storyboards, participants could provide open-ended comments.

\subsection{Participants}
An a priori power analysis determined the sample size for the between-participant personality effect, accounting for the repeated-measures structure of the mixed design. Assuming a medium effect size (\(f = 0.25\)), \(\alpha = 0.05\), and 80\% power, the analysis required 136 participants, based on \citet{ortloff_small_2025}'s meta-study of quantitative HCI effect sizes. We set the recruitment target to 140 (35 per condition). We recruited 140 U.S.-based older adults ($M = 66.31$ years; $SD = 4.78$; 98 female, 41 male, 1 other) through Prolific. Eligibility required a 98\% approval rate and at least one year of platform activity. All participants spoke English as a native or primary language. Among the participants, 118 (84.3\%) reported more than a little experience with VAs such as Google Home, Siri, and Alexa. We compensated participants at \$8.40 per hour, above the U.S. federal minimum wage.

\subsection{Data Analysis}
\label{sec:data_analysis}

Shapiro--Wilk tests indicated that the outcome distributions departed from normality, so we analyzed all seven measures with a single rank-based mixed-design model. For each outcome, we fit an Aligned Rank Transform (ART) ANOVA \cite{wobbrock_aligned_2011, elkin_aligned_2021}, $Y \sim \text{personality condition} \times \text{context} \times \text{information source} + (1 \mid \text{participant})$, with personality condition between participants and context and information source within participants, fitted with all interactions. The participant-level random intercept accounted for the four repeated measurements per participant. We used alignment appropriate to each comparison \cite{elkin_aligned_2021}. The HA--LA and HE--LE main-effect contrasts used the response aligned and ranked for personality. Simple-effect comparisons of information source within personality conditions, contexts, or personality $\times$ context combinations used ART-C alignment for the relevant combinations of factor levels. The trait-specific personality $\times$ information-source contrasts used the original ART response aligned and ranked for that interaction, averaged over context. Supplementary materials provide the corresponding model calls and contrast specifications.

RQ1 concerns the independent effects of the two layers. The conversational layer is the four-level personality factor, but because each condition varied one trait (\S \ref{sec:study_design}), only two of its six pairwise comparisons isolate a manipulated trait: HA--LA for agreeableness and HE--LE for extraversion. We therefore tested the conversational layer through these two design-defined contrasts on the personality main effect and did not test the four cross-profile pairs, such as HA--LE, which differ on both traits. Because ART contrast estimates are in aligned-rank units, we report rank-biserial correlations ($r$) on participant-level scores, averaged across each participant's four measurements, as descriptive effect sizes for these contrasts. The evidential layer is the information-source main effect, which we report together with the context main effect.

RQ2 concerns the interaction between the two layers and the variation of their effects across contexts of use. The personality $\times$ information-source term tested whether the two layers interacted, and the context $\times$ information-source, personality $\times$ context, and three-way terms tested whether the effects of either layer varied across contexts. Because the four-level interaction term does not isolate a trait, we also tested two trait-specific interaction contrasts, $(\mathrm{HA}_{\mathrm{ENV}}-\mathrm{HA}_{\mathrm{UH}})-(\mathrm{LA}_{\mathrm{ENV}}-\mathrm{LA}_{\mathrm{UH}})$ and $(\mathrm{LE}_{\mathrm{ENV}}-\mathrm{LE}_{\mathrm{UH}})-(\mathrm{HE}_{\mathrm{ENV}}-\mathrm{HE}_{\mathrm{UH}})$, as ART interaction contrasts testing differences of differences, averaged over context. When an interaction term survived adjustment for an outcome, we decomposed it into the simple effects of information source (UH vs.\ ENV) within each level of the other factor: within each personality condition, within each context, or within each personality $\times$ context cell. We did not test comparisons between cells that differ on more than one factor.

We used the Benjamini--Hochberg procedure ($p_{\mathrm{BH}}$) \cite{benjamini1995controlling} for each model term and each design-defined contrast, and across the simple effects within each decomposed interaction. We report raw and adjusted $p$-values and interpret effects that survived adjustment.

We also used three robustness or sensitivity analyses to assess central interpretations. First, we tested the HA--LA intelligence contrast for equivalence at $d=0.50$. Second, significance on one measure and non-significance on another do not establish that the effects differ. We therefore compared the HA--LA effects on empathy and intelligence and the ENV--UH effects on intelligence and empathy using bootstrap contrasts of standardized effects (Table \ref{appendix:dissociation}). Third, we assessed the LE--HE information-source interaction pattern with a covariate-adjusted ENV endpoint model using HC3-robust standard errors and a rank-based endpoint model (Table \ref{appendix:baseline_sensitivity}). These robustness analyses assessed interpretations established in the RQ analyses rather than defining additional outcome-wise decision criteria.

\section{Results}

All 140 participants completed the study and passed the attention checks; we excluded no data. Participant agreeableness (\(H(3)=2.01\), \(p=.571\); \(Mdn=4.25\)), extraversion (\(H(3)=2.51\), \(p=.474\); \(Mdn=2.88\)), and completion time (\(M = 29\) min 22 s, \(SD = 9\) min 46 s; \(H(3) = 1.97\), \(p = .580\)) did not differ across conditions.

All seven measures had good to excellent internal consistency (Cronbach's $\alpha = .889$--$.984$; McDonald's $\omega = .901$--$.984$). An exploratory factor analysis supported a one-factor solution for the reliance scale. The first eigenvalue was $3.64$, the extracted factor accounted for $66.5\%$ of the variance, and all item loadings exceeded $.65$.

\subsection{Manipulation Check}
\label{sec:manipulation_check}
We ran a manipulation-check study, recruited through Prolific, using the same interactive storyboard panel and scenarios. On each trial, participants selected which of two AI-generated responses better matched a specified personality dimension. We randomized the left--right placement of the responses and compared accuracy with chance (50\%) using one-sample $t$-tests.

We first examined whether participants could distinguish between stimuli manipulated for agreeableness and those manipulated for extraversion.  Participants distinguished agreeableness-manipulated from extraversion-manipulated stimuli ($n = 74$), $t(73) = 8.19$, $p < .001$, $d = 0.95$. We then examined whether participants could identify low versus high agreeableness and low versus high extraversion ($n = 34$ per trait). Accuracy was 97.0\% for high versus low agreeableness, $t(33) = 48.61$, $p < .001$, $d = 8.34$, and 88.0\% for high versus low extraversion, $t(33) = 11.27$, $p < .001$, $d = 1.93$. These findings indicate that participants detected the intended trait and expression-level differences in \textit{Robin}'s explanations, examined in RQ1.

\subsection{Independent Effects of the Conversational and Evidential Layers (RQ1)}
\label{sec:rq1}

RQ1 asks how each layer of \textit{Robin}'s explanations shaped perceptions on its own. We report the conversational layer through the personality main effect and the two design-defined contrasts, then the evidential layer through the information-source main effect, and close with the measures on which the two layers diverged. Section \ref{sec:rq2} tests whether these effects depended on each other and on the context of use. Tables \ref{appendix:art_anova_a}--\ref{appendix:art_anova_b} report the complete ART model.

\subsubsection{Conversational Layer: Conversational Personality}
\label{sec:rq1_personality}

\begin{figure*}[ht]
    \centering
    \includegraphics[scale=0.33]{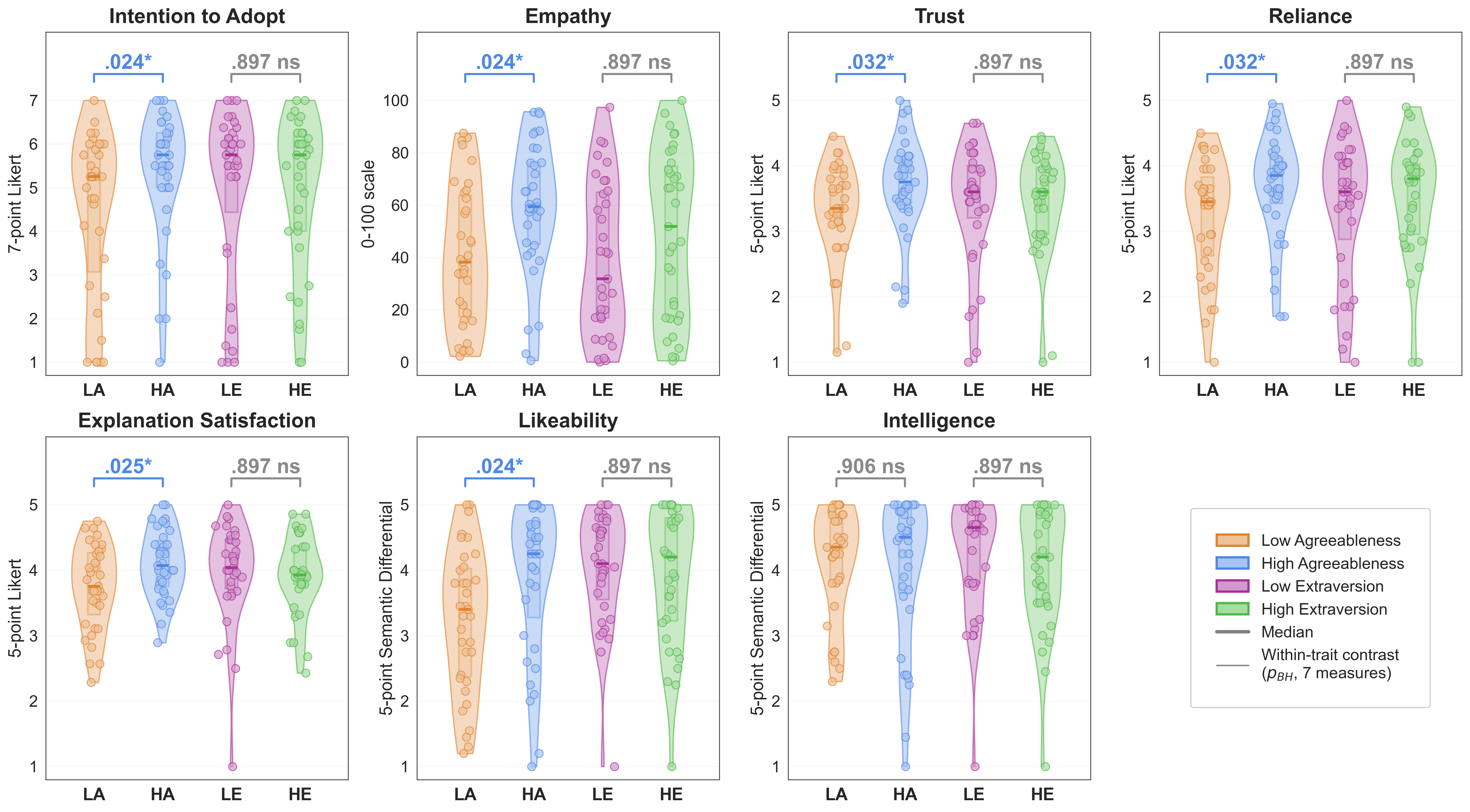}
    \caption{Outcome measures across four LLM-VA personality conditions (LA, HA, LE, HE). Violin plots show individual participant-level ratings and lines indicate medians. Brackets mark the two design-defined contrasts (LA vs.\ HA, LE vs.\ HE) and give Benjamini--Hochberg adjusted $p$-values from the ART contrasts across the seven measures, matching Table \ref{tab:within_trait_contrasts}. Colored brackets indicate a significant contrast and identify the higher-ranked condition; grey brackets are non-significant.}
    \label{fig:overview_agreeableness}
    \Description{Seven violin plots are arranged in two rows. The top row shows intention to adopt, empathy, trust, and reliance; the bottom row shows explanation satisfaction, likeability, and intelligence. Every horizontal axis orders the groups as low agreeableness (LA), high agreeableness (HA), low extraversion (LE), and high extraversion (HE). Vertical axes show ratings: 1–7 for intention to adopt, 0–100 for empathy, and 1–5 for the other outcomes. Wider parts of a violin indicate more concentrated scores; dots represent participants' average ratings and horizontal lines mark medians. Brackets compare LA with HA and LE with HE, using Benjamini–Hochberg adjusted p-values. HA exceeds LA on all outcomes except intelligence, with adjusted p-values from .024 to .032. The intelligence comparison is not significant at .906. No LE–HE comparison is significant; all adjusted p-values are .897. Empathy scores are widely spread, while intelligence ratings cluster toward the upper end of the scale. Gray brackets labeled “ns” mark comparisons with no detected difference.}
\end{figure*}

The four-level personality main effect survived adjustment for likeability only ($F(3,136)=4.27$, $p=.006$, $p_{\mathrm{BH}}=.045$, $\eta^2_p=.086$). Empathy was next ($F(3,136)=3.45$, $p=.019$, $p_{\mathrm{BH}}=.065$), and the remaining outcomes had $p_{\mathrm{BH}}\geq .154$. This omnibus term pools the two manipulated traits, so we turn to the two design-defined contrasts, which isolate agreeableness (HA vs.\ LA) and extraversion (HE vs.\ LE). We applied BH adjustment across the seven outcomes within each contrast. Cross-profile comparisons such as HA vs.\ LE do not isolate either manipulated trait and were not tested. We report the contrasts in Table \ref{tab:within_trait_contrasts} and the distributions in Fig. \ref{fig:overview_agreeableness}.

\begin{table}[!t]
\centering
\caption{Design-defined contrasts on the personality main effect of the ART model, isolating each manipulated trait. $p_{\mathrm{BH}}$ is Benjamini--Hochberg adjusted across the seven outcome measures within each contrast.}
\label{tab:within_trait_contrasts}
\begin{tabular}{@{}lrrrrc@{}}
\toprule
Measure & $t(136)$ & $p$ & $p_{\mathrm{BH}}$ & $r$ [95\% CI] & Sig. \\
\midrule
\multicolumn{6}{@{}l}{\textit{Agreeableness contrast} (HA--LA; positive $t$ and $r$ favor HA)} \\
Empathy                  & 2.60 & .010 & .024 & $.378$ $[.113,.621]$ & * \\
Likeability              & 2.92 & .004 & .024 & $.372$ $[.115,.616]$ & * \\
Trust                    & 2.30 & .023 & .032 & $.331$ $[.070,.573]$ & * \\
Intention to Adopt       & 2.60 & .010 & .024 & $.319$ $[.051,.562]$ & * \\
Reliance                 & 2.22 & .028 & .032 & $.311$ $[.046,.555]$ & * \\
Explanation Satisfaction & 2.49 & .014 & .025 & $.299$ $[.034,.545]$ & * \\
Intelligence             & 0.12 & .906 & .906 & $.051$ $[-.226,.322]$ &   \\
\addlinespace
\multicolumn{6}{@{}l}{\textit{Extraversion contrast} (LE--HE; positive $t$ and $r$ favor LE)} \\
Empathy                  & $-1.51$ & .133 & .897 & $-.192$ $[-.458,.076]$ &   \\
Likeability              & 0.17 & .864 & .897 & $.026$ $[-.250,.300]$ &   \\
Trust                    & 0.13 & .897 & .897 & $.011$ $[-.262,.286]$ &   \\
Intention to Adopt       & 0.27 & .786 & .897 & $.033$ $[-.249,.304]$ &   \\
Reliance                 & $-0.28$ & .779 & .897 & $-.039$ $[-.313,.238]$ &   \\
Explanation Satisfaction & 0.57 & .571 & .897 & $.092$ $[-.181,.365]$ &   \\
Intelligence             & 0.45 & .650 & .897 & $.084$ $[-.184,.353]$ &   \\
\bottomrule
\end{tabular}
\vspace{2mm}

\footnotesize \textit{Note.} $t$ = ART-C contrast on the personality main effect of the mixed model ($n = 35$ per condition; Kenward--Roger $df$). $r$ = rank-biserial correlation on participant-level scores averaged across each participant's four repeated measurements, with bootstrap percentile 95\% CI, reported as a descriptive effect size. HA ranked higher than LA on all seven measures; directions were mixed for LE--HE. The seven extraversion $p_{\mathrm{BH}}$ values are identical because the step-up procedure caps each adjusted value at the largest one in the family when no test approaches the threshold. Significance codes refer to $p_{\mathrm{BH}}$: $^{*}p_{\mathrm{BH}}<.05$.

\end{table}

Explanations delivered by HA \textit{Robin} received higher ratings than those delivered by LA \textit{Robin} on six measures: empathy, likeability, trust, intention to adopt, reliance, and explanation satisfaction (all $p_{\mathrm{BH}} \leq .032$; $r = .299$--$.378$). We detected no HA--LA difference in perceived intelligence ($p_{\mathrm{BH}} = .906$, $r = .051$). We read this as a profile-level shift: HA ranked higher than LA on all seven measures, the six detected contrasts had similarly sized effects, and empathy and likeability showed the largest differences ($r = .378$ and $.372$).

By contrast, we did not detect a source-averaged LE--HE difference on any measure (all $p_{\mathrm{BH}} = .897$). Directions were mixed: HE was numerically higher on empathy and reliance, and LE on the remaining five. Intelligence responses were compressed at the top of the five-point scale: condition medians ranged from 4.20 to 4.65, 52.1\% of item responses used the maximum category, 39.1\% of the 560 composite scores (one per participant per measurement occasion) were at the maximum, and 17.9\% of participant-level composites averaged across the four occasions were at the maximum.

For the central HA--LA intelligence contrast, a robustness equivalence test at $d=0.50$ was significant (TOST $p=.040$; $d=-0.07$, 95\% CI $[-0.54,0.39]$). This result rules out effects at that bound but remains compatible with smaller effects.

\subsubsection{Evidential Layer: Information Source}
\label{sec:rq1_source}

The evidential layer varied within participants, so we tested it through the information-source main effect of the same model, which averages over the four personality conditions and both contexts. Table \ref{tab:source_main_effect} reports this effect together with the participant-level distributions for each information source. ENV explanations received higher ratings than UH explanations on five measures: explanation satisfaction, reliance, trust, intelligence, and intention to adopt (all $p_{\mathrm{BH}}\leq .031$; $\eta^2_p=.013$--$.044$). The corresponding raw-scale differences were small, from $+0.10$ points for trust to $+0.25$ for intention to adopt. We detected no information-source main effect on empathy or likeability (both $p_{\mathrm{BH}}=.122$).

\begin{table*}[t]
\centering
\caption{Information-source main effect from the ART model, with participant-level descriptive statistics for each source. $p_{\mathrm{BH}}$ is Benjamini--Hochberg adjusted across the seven outcome measures within the effect term.}
\label{tab:source_main_effect}
\begin{tabular}{@{}lccrrrrrc@{}}
\toprule
& \multicolumn{2}{c}{Median (IQR)} & & & & & & \\
\cmidrule(lr){2-3}
Measure & UH & ENV & $\Delta$ & $F(1,408)$ & $p$ & $p_{\mathrm{BH}}$ & $\eta^2_p$ & Sig. \\
\midrule
Empathy                   & 50.43 (52.45)   & 48.14 (52.20)   & $-1.64$ & 2.40 & .122 & .122 & .006 &   \\
Likeability               & 4.00 (2.00)     & 4.00 (1.52)     & $+0.13$ & 2.47 & .117 & .122 & .006 &   \\
Trust                     & 3.60 (1.00)     & 3.60 (0.90)     & $+0.10$ & 7.63 & .006 & .014 & .018 & * \\
Intention to Adopt        & 5.50 (2.00)     & 6.00 (2.25)     & $+0.25$ & 5.25 & .022 & .031 & .013 & * \\
Reliance                  & 3.60 (1.20)     & 3.80 (1.00)     & $+0.17$ & 17.95 & $<.001$ & $<.001$ & .042 & *** \\
Explanation Satisfaction  & 4.00 (0.93)     & 4.00 (0.73)     & $+0.18$ & 18.70 & $<.001$ & $<.001$ & .044 & *** \\
Intelligence              & 4.40 (1.45)     & 4.50 (1.20)     & $+0.17$ & 6.54 & .011 & .019 & .016 & * \\
\bottomrule
\end{tabular}
\vspace{2mm}

\footnotesize \textit{Note.} UH = Conversational User History; ENV = Real-time Environmental Data. Descriptives are participant-level scores averaged across the two contexts within each information source ($N = 140$); $\Delta$ is the mean raw-scale difference (ENV $-$ UH). Empathy uses a 101-point slider, intention to adopt a 7-point Likert scale, and the remaining five measures 5-point scales, so $\Delta$ is not comparable across rows. Significance codes refer to $p_{\mathrm{BH}}$: $^{***}p_{\mathrm{BH}}<.001$, $^{**}p_{\mathrm{BH}}<.01$, $^{*}p_{\mathrm{BH}}<.05$.
\end{table*}

The two layers, therefore, showed different patterns across outcomes. Information source affected perceived intelligence but not empathy or likeability, whereas agreeableness affected empathy and likeability but not perceived intelligence. We detected no context main effect after adjustment; the smallest adjusted $p$-value was for intelligence ($F(1,408)=6.85$, $p=.009$, $p_{\mathrm{BH}}=.064$). However, the information-source effect varied by context, as Section \ref{sec:context_explanation} reports.

\subsubsection{Where the Two Layers Diverged: Empathy and Perceived Intelligence}
\label{sec:rq1_dissociation}

Empathy and intelligence responded to different layers (Fig. \ref{fig:empathy_intelligence}). The HA--LA contrast differed on empathy ($p_{\mathrm{BH}}=.024$), whereas we detected no information-source effect on empathy ($p_{\mathrm{BH}}=.122$). For intelligence, we detected no personality effect (personality main effect $p_{\mathrm{BH}}=.953$; HA--LA $p_{\mathrm{BH}}=.906$), but detected an information-source effect ($p_{\mathrm{BH}}=.019$) despite ceiling compression. Direct cross-measure tests supported this differential sensitivity (Table \ref{appendix:dissociation}): the HA--LA effect was larger on empathy than on intelligence ($\Delta d=0.78$, 95\% CI $[0.23,1.39]$, bootstrap $p=.006$), and the ENV--UH effect was larger on intelligence than on empathy ($\Delta=0.24$ SD, 95\% CI $[0.13,0.35]$, bootstrap $p<.001$).

\begin{framed}
\noindent\textbf{\textit{Summary.}} Within the conversational layer, HA received higher ratings than LA on six of seven measures (all $p_{\mathrm{BH}} \leq .032$; $r = .299$--$.378$), with the largest differences on empathy and likeability and no detected difference in perceived intelligence. We detected no LE--HE differences. Within the evidential layer, ENV received higher ratings than UH on explanation satisfaction, reliance, trust, intelligence, and intention to adopt (all $p_{\mathrm{BH}} \leq .031$), with no detected difference in empathy or likeability. Agreeableness differentiated empathy but not intelligence, whereas information source differentiated intelligence but not empathy.
\end{framed}

\subsection{Interaction Between the Layers and Variation Across Contexts of Use (RQ2)}
\label{sec:rq2}

RQ2 asks whether the two layers interacted, that is, whether the information-source effect depended on \textit{Robin}'s conversational personality, and whether the effects of either layer varied between routine reminders and emergency alerts. The analysis comprised the interaction terms of the full 16-cell model, the two trait-specific interaction contrasts, the simple-effect decompositions of corrected interactions, and robustness checks for the LE--HE pattern. Tables \ref{appendix:art_anova_a}--\ref{appendix:art_anova_b} report the complete ART model; the supplementary materials report descriptive statistics and all simple effects.

\subsubsection{Personality $\times$ Information Source Interaction}
\label{sec:personality_source}
The four-level personality $\times$ information-source term showed heterogeneity in source effects across profiles for trust, reliance, explanation satisfaction, and intention to adopt (all $p_{\mathrm{BH}} \leq .022$). Likeability ($p=.188$, $p_{\mathrm{BH}}=.264$), empathy, and intelligence (both $p_{\mathrm{BH}}=.349$) did not survive adjustment. We decomposed the four corrected interactions into the UH-to-ENV simple effect within each personality condition. LE increased on all four outcomes (all $p_{\mathrm{BH}} \leq .001$); for example, trust increased by $0.31$ points and intention to adopt by $0.63$ points. LA increased on reliance ($+0.22$, $p_{\mathrm{BH}}=.007$) and explanation satisfaction ($+0.22$, $p_{\mathrm{BH}}=.013$) but not on trust or intention to adopt. HA and HE did not differ between information sources on any of the four outcomes (all $p_{\mathrm{BH}} \geq .168$).

To determine whether the information-source effect differed between the two extraversion profiles, we directly tested $(\mathrm{LE}_{\mathrm{ENV}}-\mathrm{LE}_{\mathrm{UH}})-(\mathrm{HE}_{\mathrm{ENV}}-\mathrm{HE}_{\mathrm{UH}})$, averaged over context. The extraversion $\times$ information-source contrast showed that LE increased more than HE on the same four outcomes: trust ($t(408)=4.27$, $p_{\mathrm{BH}}<.001$), explanation satisfaction ($t(408)=3.14$, $p_{\mathrm{BH}}=.004$), reliance ($t(408)=3.13$, $p_{\mathrm{BH}}=.004$), and intention to adopt ($t(408)=3.09$, $p_{\mathrm{BH}}=.004$). For example, trust increased by $0.31$ points in LE and changed by $-0.03$ in HE, and intention to adopt increased by $0.63$ in LE and $0.02$ in HE. Likeability did not survive adjustment ($t(408)=1.91$, $p=.057$, $p_{\mathrm{BH}}=.079$), and we detected no contrast for empathy or intelligence (both $p_{\mathrm{BH}} \geq .150$). We detected no corresponding agreeableness $\times$ information-source contrast (all $p_{\mathrm{BH}} \geq .146$): the UH-to-ENV changes in LA were not distinguishable from those in HA, and the HA--LA differences reported in RQ1 were similar under both information sources.

A covariate-adjusted endpoint sensitivity analysis retained the LE--HE pattern on the four outcomes after adjustment for UH ratings ($b=0.253$--$0.582$, $p_{\mathrm{BH}}<.001$--$.014$; Table \ref{appendix:baseline_sensitivity}), and additionally reached adjusted significance for likeability ($b=0.331$, $p_{\mathrm{BH}}=.014$). A rank-based endpoint analysis yielded the same pattern. Because UH was a counterbalanced condition rather than a pretreatment baseline, these analyses address scale headroom and serve as robustness checks rather than alternative outcome-wise decision criteria.

\subsubsection{Context $\times$ Information Source Interaction}
\label{sec:context_explanation}
Context interacted with the information source on six of seven measures, all except empathy (all $p_{\mathrm{BH}} \leq .002$; empathy $p_{\mathrm{BH}}=.089$).

\begin{figure*}[ht]
    \centering
    \includegraphics[scale=0.10]{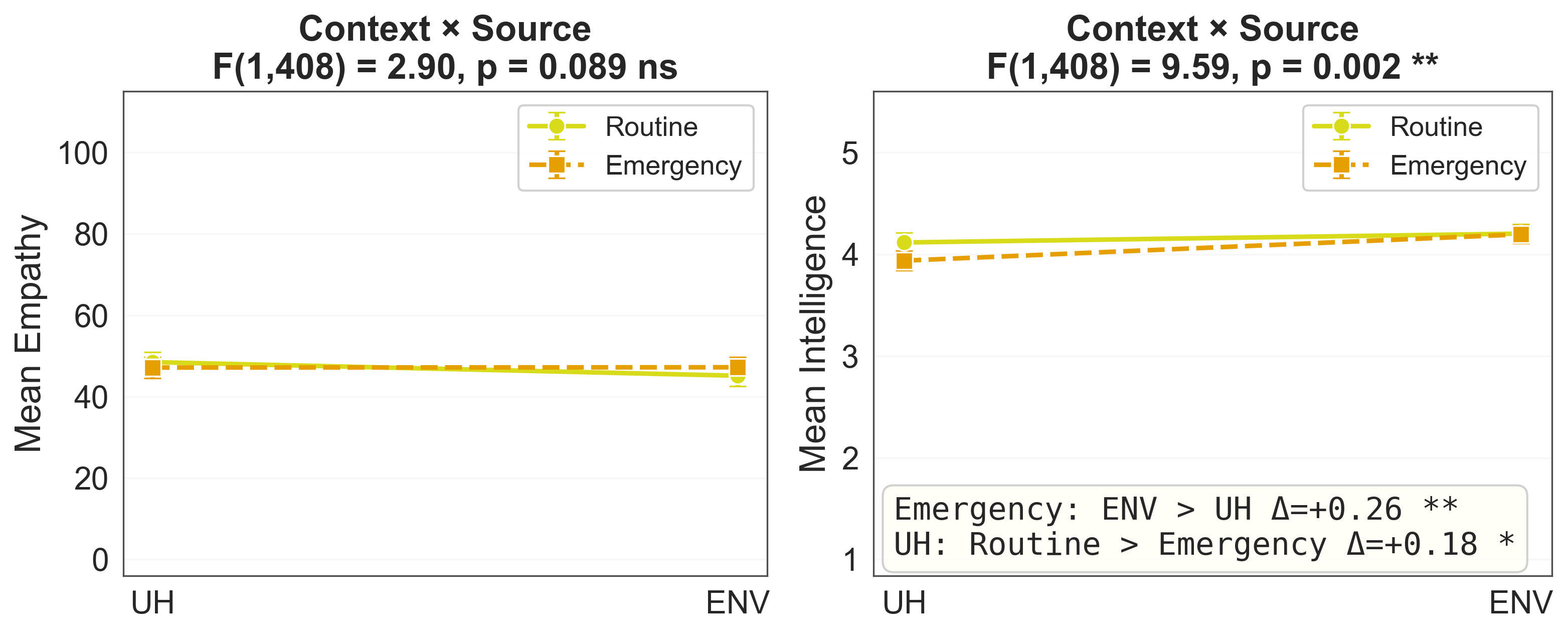}
    \caption{Context $\times$ Information Source interaction for empathy (left) and perceived intelligence (right). No context, information-source, or interaction effect was detected for empathy (all $p_{\mathrm{BH}} \geq .089$). No personality effect was detected for intelligence ($p_{\mathrm{BH}} = .953$), whereas its Context $\times$ Information Source interaction was significant ($F(1,408) = 9.59$, $p_{\mathrm{BH}} = .002$), with ENV explanations producing higher ratings than UH primarily in the emergency context ($\Delta = +0.26$, $p_{\mathrm{BH}} = .002$). Intelligence scores were compressed at the top of the scale.}
    \label{fig:empathy_intelligence}
    \Description{Two line plots compare conversational user history (UH), on the left of each horizontal axis, with real-time environmental data (ENV), on the right. Solid lines with circles represent routine reminders; dashed lines with squares represent emergency alerts. Points show means with vertical error bars. The left plot uses the 0–100 empathy scale. Both context lines remain near the middle of the scale and change little between sources; no context, source, or interaction effect is statistically detected. The right plot uses the 1–5 intelligence scale. The routine line remains near 4.1, while the emergency line rises from about 3.9 under UH to about 4.2 under ENV, converging with the routine line. An annotation identifies a 0.26-point ENV advantage during emergencies. Another identifies a 0.18-point routine-over-emergency difference under UH. The context-by-source interaction is significant for intelligence, with adjusted p = .002, but not empathy, with adjusted p = .089.}
\end{figure*}

We decomposed the six corrected interactions into the UH-to-ENV simple effect within each context. UH and ENV ratings did not differ in routine reminders on any of the six measures ($p_{\mathrm{BH}}=.148$--$.963$). In emergency alerts, ENV received higher ratings than UH on intention to adopt ($+0.55$ points), explanation satisfaction ($+0.39$), reliance ($+0.36$), likeability ($+0.29$), intelligence ($+0.26$), and trust ($+0.25$; all $p_{\mathrm{BH}}\leq .003$). The RQ1 information-source main effect therefore came primarily from the emergency context rather than uniformly across contexts. Likeability, which showed no information-source main effect, followed the same pattern: ENV exceeded UH only in emergencies. Fig. \ref{fig:empathy_intelligence} shows the interaction for the two measures on which the layers diverged in RQ1: empathy showed no context, information-source, or interaction effect (all $p_{\mathrm{BH}} \geq .089$), whereas intelligence showed a context $\times$ information-source interaction ($p_{\mathrm{BH}}=.002$) with the ENV advantage confined to emergencies.

\subsubsection{Personality $\times$ Context and Three-Way Interactions}
\label{sec:three_way}

\begin{figure*}[ht]
    \centering
    \includegraphics[scale=0.30]{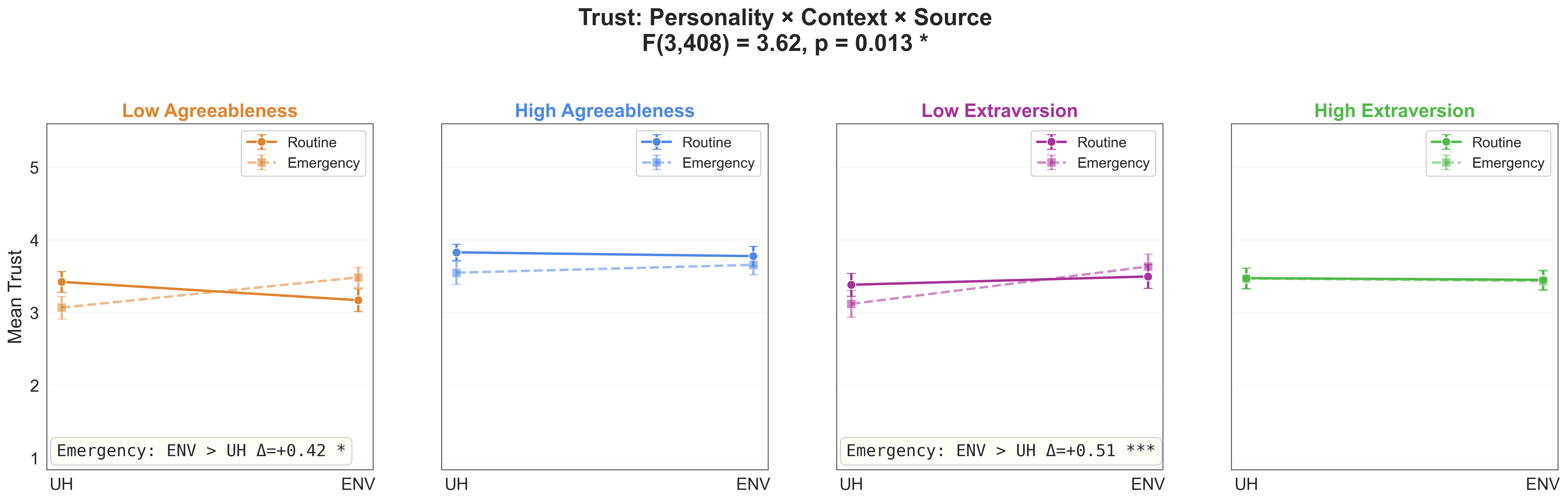}
    \caption{Personality $\times$ Context $\times$ Information Source interaction for trust ($F(3,408)=3.62$, $p=.013$, $p_{\mathrm{BH}}=.046$). In the simple-effect decomposition, ENV exceeded UH in the LE ($\Delta=+0.51$, $p_{\mathrm{BH}}<.001$) and LA ($\Delta=+0.42$, $p_{\mathrm{BH}}<.001$) emergency cells; HA and HE remained stable. Error bars show standard errors.}
    \label{fig:trust}
    \Description{Four line plots run from left to right: low agreeableness, high agreeableness, low extraversion, and high extraversion. Each compares conversational user history (UH) with real-time environmental data (ENV) on the horizontal axis and mean trust on a 1–5 vertical scale. Solid lines with circles show routine reminders; dashed lines with squares show emergency alerts. Error bars represent standard errors. In low agreeableness, emergency trust rises from UH to ENV while routine trust declines, so the lines cross. In low extraversion, emergency trust rises more steeply than routine trust. The annotated emergency increases are 0.42 points for low agreeableness and 0.51 points for low extraversion; both are significant according to the adjusted tests reported in the caption. High agreeableness shows nearly parallel, gently declining lines, and high extraversion shows overlapping, nearly flat lines. No source difference is detected in the other personality-by-context combinations. The overall three-way interaction has adjusted p = .046.}
\end{figure*}

No personality $\times$ context interaction survived adjustment; the strongest was intention to adopt ($F(3,408)=3.03$, $p=.029$, $p_{\mathrm{BH}}=.204$). The conversational-layer effects reported in RQ1 therefore did not differ detectably between routine reminders and emergency alerts. Three-way interactions survived for trust and likeability ($p_{\mathrm{BH}}=.046$ for both). We decomposed each into the UH-to-ENV simple effect within the eight personality $\times$ context cells. For trust, ENV exceeded UH only in the LE ($+0.51$, $p_{\mathrm{BH}}<.001$) and LA ($+0.42$, $p_{\mathrm{BH}}<.001$) emergency cells; HA and HE did not differ in either context (Fig. \ref{fig:trust}). For likeability, ENV exceeded UH in the HA emergency cell ($+0.51$, $p_{\mathrm{BH}}=.049$), with the LA emergency cell at the threshold ($+0.49$, $p_{\mathrm{BH}}=.052$); no routine cell differed (all $p_{\mathrm{BH}} \geq .089$). These decompositions located the ENV advantage in particular emergency cells rather than a consistent ordering across conditions, so we do not develop broader claims from them.

\begin{framed}
\noindent\textbf{\textit{Summary.}} The two layers interacted: personality moderated the information-source effect on trust, reliance, explanation satisfaction, and intention to adopt. LE ratings increased with ENV on all four measures, LA on two, and HA and HE on none, and the extraversion-specific contrast confirmed larger LE than HE gains on the four, whereas HA--LA differences were similar across information sources. Context qualified the evidential layer but not the conversational layer: UH and ENV were comparable in routine reminders, whereas ENV received higher ratings on six measures in emergency alerts, and no personality $\times$ context interaction survived adjustment. The two three-way interactions located the ENV advantage in particular emergency cells.
\end{framed}

\subsection{Qualitative Perceptions}

After completing all 16 storyboards and quantitative measures, participants were invited to provide an optional open-ended comment about Robin. Of the 140 participants, 85 provided a response (LA: 23; HA: 19; LE: 23; HE: 20). We use these comments descriptively to contextualize the quantitative findings rather than as a separate qualitative analysis. 

Responses to the agreeableness manipulation were particularly distinct. Participants in the LA condition repeatedly described Robin as ``rude,'' ``sarcastic,'' ``condescending,'' ``snarky,'' or ``nagging,'' with several explicitly objecting to being spoken to in a directive or judgmental manner. One participant noted that while Robin's suggestions and explanations were adequate, its delivery felt ``negative and unsupportive.'' In contrast, participants exposed to HA explanations more often described Robin as ``pleasant,'' ``comforting,'' ``thoughtful,'' and ``helpful,'' including one participant who felt that Robin's phrasing did not make Arthur ``feel less capable of managing things.'' These perceptions are consistent with the higher empathy, likeability, trust, reliance, explanation satisfaction, and adoption ratings observed for HA over LA.

Comments also reflected differences in how participants interpreted evidential grounding. Several participants explicitly valued explanations based on current sensor information because they appeared more concrete or verifiable. For example, one participant preferred the version that ``used verifiable data input,'' while another found sensor-grounded explanations more acceptable because Robin could ``actually quantify the situation or problem.'' This preference was particularly apparent when participants discussed safety and emergency scenarios, although some simultaneously raised concerns about privacy and excessive monitoring. Together, these comments reinforce the quantitative finding that environmental evidence was not uniformly preferred, but became especially valuable when explanations concerned immediate or potentially consequential situations.

Finally, comments about extraversion were more mixed. HE Robin was variously described as ``friendly,'' ``supportive,'' and ``upbeat,'' but also at times as ``naggy'' or overly expressive. LE Robin was described as ``calm,'' ``analytical,'' and focused, while others found it ``curt,'' ``machine like,'' or insufficiently warm. This heterogeneity is consistent with the absence of an overall HE--LE preference and suggests that the reception of extraverted or reserved delivery depended more strongly on what information accompanied it.

\section{Discussion}

We conducted a mixed-design experiment to examine how older adults perceived explanations from an LLM-VA across different contexts of use and with varying sources of information and conversational personalities. Our findings suggest that conversational explanations for older adults should be understood as comprising two complementary design dimensions: an evidential layer, which concerns what information an explanation references, and a conversational layer, which concerns how that information is communicated. Across both routine reminders and emergency alerts, perceptions of trust, reliance, explanation satisfaction, and intention to adopt responded to both layers and their interaction. Our results show that explanations based on the same information source can be received differently by older adults, depending on the conversational personality of the LLM-VA. We therefore did not find any combination of conversational and evidential layers for LLM-VA explanations that older adults perceived as best across all contexts. 

\subsection{Distinct Effects of Conversational Delivery and Evidential Grounding}
\label{sec:empathy-intel}

Our study design separated how \textit{Robin} expressed conversational personality from the information source cited in its explanations. This allowed us to examine the persona of the conversational layer \cite{mathur2026explanations} through \textit{conversational personality} and the evidential grounding of explanations through \textit{information sources}. We then tested whether these two layers shaped the same perceptions among older adults. 

The contrast between the high- and low-agreeableness conditions was largest for empathy and likeability, and we detected no difference in perceived intelligence. By contrast, information source differentiated perceived intelligence, whereas we detected no difference in empathy. Direct cross-measure tests supported this difference. The agreeableness effect was larger for empathy than for intelligence ($\Delta d=0.78$, bootstrap $p=.006$). The effect of changing from Conversational User History to Real-time Environmental Data was larger for intelligence than for empathy ($\Delta=0.24$ SD, bootstrap $p<.001$; Table \ref{appendix:dissociation}). Descriptively, the extraversion conditions showed a reverse directional pattern, with low-extraversion receiving higher perceived-intelligence ratings and lower empathy ratings than high-extraversion.

This difference corresponds to the warmth and competence dimensions of the Stereotype Content Model (SCM) \cite{fiske_universal_2007}, which also organize impressions of AI systems \cite{mckee_humans_2023}. Prior work similarly shows that conversational cues can alter warmth without changing competence judgments \cite{heppner_conveying_2024} and that people can prefer high-warmth AI systems even when competence is lower \cite{gilad_effects_2021}. Our findings extend this distinction to LLM-VA explanations for older adults by showing that perceived empathy and intelligence respond to different aspects of the explanation.

The conversational personality and information source were not fully independent. The effect of information source varied across conversational personality profiles for trust, reliance, explanation satisfaction, and intention to adopt. Our findings, therefore, support evaluating conversational personality and information source separately while also testing their interaction.

\subsection{Agreeableness and Interpersonal Expectations in the Conversational Layer}
\label{sec:differential_role}

We manipulated the conversational layer through four single-trait conversational personality conditions. TMK expressed agreeableness and extraversion through the language of \textit{Robin}'s explanations (\S \ref{sec:design_personality}). The manipulation checks supported intended-trait recognition. The contrast between the high- and low-agreeableness conditions, therefore, isolates agreeableness at the complete conversational profile level.

The high-agreeableness condition received higher ratings than the low-agreeableness condition on empathy, likeability, trust, reliance, explanation satisfaction, and intention to adopt after BH adjustment ($p_{\mathrm{BH}} \leq .033$, $r=.299$--$.378$). We detected no HA--LA difference in perceived intelligence. The six detected contrasts had similar effect sizes, and we detected no agreeableness $\times$ information-source contrast. This pattern supports the profile-level shift across social and evaluative measures reported in RQ1 and appeared under both the Conversational User History and Real-time Environmental Data explanations.

Within aging in place environments, older adults often interact with voice assistants over extended periods as part of their everyday routines rather than as one-time information tools \cite{zubatiy2021empowering, pradhan2020use}. As discussed in Section 2, prior work suggests that these interactions are evaluated for more than functional utility, and also through interpersonal expectations of warmth, respect, and competence \cite{khadpe_conceptual_2020, desai_examining_2024}. Our findings extend this perspective by demonstrating that the conversational delivery of an explanation substantially shapes these relational perceptions even when the evidential grounding remains constant.

Politeness theory and older adults' relational expectations offer one account for this pattern. \citet{brown_politeness_1987} argues that interlocutors manage each other's ``face needs''---the desire for respect, liking, and approval---through linguistic strategies. The high-agreeableness explanations used politeness markers, hedges, and cooperative phrasing, whereas the low-agreeableness explanations were more directive and critical. Agreeableness includes both compassion and politeness \cite{deyoung_between_2007}, and both were expressed by \textit{Robin}.

Operationalization of personality at the profile level allows us to test the conversational profile \cite{volkel_user_2022, rahman_vibe_2025}. Related work shows that politeness strategies influence comfort, trust, and liking in interactions with CAs and VAs \cite{bowman_exploring_2024, cox_polite_2026, kim_should_2025, hu_polite_2022}.

The interpretation is also consistent with research showing that older adults value warmth, respect, and rapport alongside functional utility and evaluate VAs through interpersonal lenses \cite{bickmore_its_2005, morrow_framework_2021, zhong_effects_2022, wong_voice_2024, chin2024like, Hu_Desai_Lundy_Chin_2025}. The low-agreeableness condition may have violated these interpersonal expectations, consistent with evidence that users penalize disagreeable conversational agents \cite{volkel_examining_2021}. 

\subsection{Evidential Layer Across Contexts of Use}
\label{sec:evidential_layer}
We operationalize the evidential layer of the explanation by varying its information source. Our findings show that the effect of the information source of an explanation depended on the context of use. Context $\times$ information-source interactions survived BH adjustment on six of seven measures (all $p_{\mathrm{BH}} \leq .002$). The simple effects showed no differences between Conversational User History and Real-time Environmental Data for routine reminders, whereas Real-time Environmental Data received higher ratings on all six affected measures for emergency alerts (all $p_{\mathrm{BH}} \leq .003$). Real-time Environmental Data was therefore not uniformly preferred. Its advantage only emerged when the context of use in our study involved urgency.

This context-dependent pattern follows the view that explanations are contrastive and selective answers to an implicit ``why'' question \cite{miller2019explanation}. Also consistent with the HCXAI perspective that explanations should satisfy users' informational needs, our findings suggest that the usefulness of an evidential source depends on the context in which an explanation is delivered. We found that older adults need context-appropriate answers to implicit ``why'' questions rather than complete causal accounts. In routine contexts, both explanations, relying on either information source, can answer ``Why remind me now?'' by referencing user preferences or current environmental conditions. In emergencies, real-time environmental explanations answer ``Why are you alerting me?'' with observable, real-time sensor data. 


Prior work with older adults similarly found that explanation preferences depend on content, tone, framing, and perceived risk and that explanations are understood as parts of an interactive exchange \cite{mathur2026sometimes}. Our results add quantitative evidence that the same two information sources can be evaluated differently across contexts of use. This also speaks to prior work showing that explanatory needs vary with application criticality \cite{lim_why_2009, lim_assessing_2009} and with calls to situate explanations in their social and environmental context \cite{ehsan2021expanding}. An LLM-VA for older adults should therefore select an information source based on the context in which it is explaining an action.

\subsection{Interaction Between Personality and Information Source}
\label{sec:layers_meet}

Extraversion did not produce a source-averaged difference in evaluations in RQ1, but it moderated the change in perceptions with information sources in RQ2. The four-level personality $\times$ information-source interaction was detected for trust, reliance, explanation satisfaction, and intention to adopt. The trait-specific contrast showed larger increases from Conversational User History to Real-time Environmental Data in the low- than high-extraversion condition for these outcomes. We detected no corresponding agreeableness $\times$ information-source contrast.

One possible interpretation follows from how the extraversion conditions were expressed. The high-extraversion condition used socially expressive language, whereas the low-extraversion condition used terse, matter-of-fact language. Extraversion-like descriptions of VAs tend to cluster around a social-entertaining dimension, while agreeableness-related descriptions occur across several positive dimensions \cite{volkel_developing_2020}. Extraversion may therefore depend more on whether users approach the VA as a social companion or as an information tool. Older adults can move between these views of a VA \cite{pradhan_phantom_2019}, and in-home assistants can be evaluated primarily as helpful services \cite{gambino_building_2020}.

Extraverted language is typically more verbose and contains more positive-emotion terms, whereas low-extraversion language is more concise and concrete \cite{pennebaker_linguistic_2001, gill2019taking}. In the Real-time Environmental Data condition, the concise low-extraversion profile may have presented the sensor data as direct evidence for the assistant's action. This interpretation is consistent with work on information-processing demands for older adults \cite{wingfield_speech_2003, stine_how_1990}, transparency in personality-adaptive CAs \cite{ahmad_designing_2022}, and assertive AI in high-risk interactions \cite{zhou_trusting_2019}. Other studies also report context-dependent effects of extraversion, asymmetric matching, and differences between stated preferences and engagement \cite{ahmad_extrabot_2020, Snyder2023OneVoiceFitsAll, volkel_user_2022}.

\subsection{Design Implications}

Our findings show that for older adults, conversational personality and information sources are not interchangeable routes to a better explanation. They shape older adults' perceptions of LLM-VA explanations both separately and in combination across contexts of use. We suggest four design implications for designers of LLM-VA explanations for older adults. These implications follow the human-centered perspective introduced in Section 2. They address both the evidential grounding of explanations (what information is communicated) and their conversational delivery (how that information is communicated), recognizing that both dimensions shape older adults' understanding of AI behavior.

\subsubsection{Give Conversational Delivery and Evidence Separate Controls}

An explanation can sound caring to an older adult without appearing more intelligent, or cite strong evidence without appearing more empathic. In our study, conversational personality differentiated empathy, whereas information source differentiated perceived intelligence (\S \ref{sec:empathy-intel}). Treating both responses as a single measure of explanation quality would hide this distinction. For older adults aging in place, this distinction is particularly important because conversational assistants are expected to support both relational and functional needs over extended periods of use as older adults' cognitive and physical needs evolve \cite{pradhan_phantom_2019, shandilya2022understanding}. 

Designers should therefore keep conversational delivery and evidence selection separately tunable. This separation allows them to adjust how an explanation is worded and which information it cites while checking relational perceptions, such as empathy, alongside competence perceptions, such as perceived intelligence.

\begin{quote}
\noindent\textbf{DI1: Separate but Coordinated Controls.} Tune conversational delivery and evidential grounding separately, then evaluate how each affects older adults' perceptions.
\end{quote}

\subsubsection{Build Warmth Into Explanations for Older Adults}

The high-agreeableness condition received higher ratings than the low-agreeableness condition on empathy, likeability, trust, reliance, explanation satisfaction, and intention to adopt under both information sources. For older adults, the wording of an explanation communicates respect, concern, and social responsiveness in addition to justification of system behavior. These qualities are central to older adults' evaluations of VAs and relational agents \cite{bickmore_its_2005, morrow_framework_2021, chin2024like}.

For older adults, explanations are not simply informational justifications but conversational exchanges within an ongoing relationship with a voice assistant \cite{mathur2026sometimes}. Designers should therefore treat warmth as part of explanation design rather than as an optional conversational afterthought. Cooperative framing, nonjudgmental wording, and acknowledgment of user agency can help explanations communicate both justification and interpersonal respect, even when an explanation is brief.

\begin{quote}
\noindent\textbf{DI2: Relational Warmth.} Build warmth into explanations for older adults through respectful, cooperative, and nonjudgmental language, including when the message must be concise.
\end{quote}

\subsubsection{Match Evidence to the Stakes of the Situation}

Information source mattered most in emergency alerts. In routine reminders, explanations of Conversational User History and Real-time Environmental Data received comparable ratings. In emergencies, explanations grounded in current sensor data received higher ratings and connected the alert to observable conditions. This pattern argues against using one information source across all interactions.

Within aging in place environments, older adults often interpret AI recommendations alongside familiar routines, household artifacts, and observable environmental cues \cite{pradhan2023towards, mathur2025feels}. Our findings suggest that explanations should therefore reference the form of evidence that is most meaningful for the situation. Conversational history can reinforce continuity and long-term understanding during routine interactions, whereas observable environmental evidence can more effectively justify urgent alerts by helping older adults immediately relate the assistant's recommendation to their surroundings. The system should therefore select and present the source that explains why the action is needed in that context and why it is needed at a given moment \cite{miller2019explanation, mathur2026sometimes, ehsan2021expanding}.

\begin{quote}
\noindent\textbf{DI3: Context-Appropriate Evidence.} Use current, observable evidence to justify time-sensitive alerts, and retain conversational history as an option for routine interactions.
\end{quote}

\subsubsection{Test Personality--Information source Combinations}
The low-extraversion condition did not differ from the high-extraversion condition overall, but its ratings increased more when explanations shifted from Conversational User History to Real-time Environmental Data. A reserved delivery style may therefore be received differently depending on the evidence supporting it.

Rather than tuning conversational personality or information sources independently, designers of LLM-based VAs for older adults should evaluate these dimensions together as combinations rather than selecting each in isolation. A concise, matter-of-fact response may work well when it cites concrete evidence, whereas conversational expressiveness cannot substitute for relevant grounding. The result supports testing combinations but does not prescribe a default delivery style.

\begin{quote}
\noindent\textbf{DI4: Personality--Information Source Fit.} Test concise, reserved delivery with concrete evidential grounding, and compare personality--information source combinations before deployment.
\end{quote}

\section{Limitations and Future Work}
\label{sec:limitations}
The study has six limitations. First, since our study was conducted on Prolific, older adult participants were likely to be more technologically experienced, comfortable with online studies and familiar with digital interfaces than many older adults who may use LLM-VAs in aging in place contexts. As a result, our findings may not fully capture how older adults with limited or no prior experience with VAs, LLMs or digital technologies would interpret and respond to conversational AI explanations. Future work should examine these explanation designs with older adults who are less technologically experienced, including those with limited access to digital tools, lower confidence using AI systems or greater reliance on caregivers and assistive technologies in their daily lives.

Second, our controlled study design involving interactive storyboards helped keep the conversational stimuli consistent across conditions while grounding them in vivid, everyday scenarios \cite{truong_storyboarding_2006}. However, storyboards are no substitute for in situ use, and our measures therefore capture participants' evaluations of the presented interactions rather than their observed behavior. Future deployments and longitudinal studies should test whether these patterns persist in everyday use.

Third, we examined high and low expressions of agreeableness and extraversion but did not include a neutral-personality condition or fully cross the two traits. Our findings, therefore, apply to the profiles studied here rather than the broader space of possible LLM-VA personalities.

Fourth, personality was manipulated at the profile level through related differences in tone, politeness, directness, affect, and verbosity. Manipulation checks support recovery of the intended traits, but the design cannot identify which linguistic features drove the effects. Component-level ablation studies could examine these mechanisms more directly.

Fifth, perceived-intelligence ratings showed a substantial ceiling effect, with 52.1\% of item responses and 39.1\% of per-occasion composite scores at the maximum. Although equivalence tests bounded the personality null at $d = 0.50$, smaller effects may have been difficult to detect. We also used a single synthetic voice with fixed affect and pitch to isolate textual personality expression, leaving interactions with prosody and vocal affect for future work.

Sixth, we manipulated the evidence source in Robin’s explanations, but not whether that evidence was correct. Environmental explanations always reported accurate sensor readings, so evidential grounding should not be confused with evidential correctness. Because reliance increased with environmental grounding, we cannot determine whether this would remain well calibrated when sensor evidence is stale, uncertain, or incorrect \cite{lammert2024humans, scharowski2023exploring, vasconcelos2023explanations}. This is especially important in emergency contexts, where the environmental advantage was strongest and incorrect evidence could be most consequential. Future work should directly manipulate evidence quality and examine how uncertainty should be communicated across different conversational styles.

\section{Conclusion}

Older adults in our study responded differently to conversational personality and information sources. The high-agreeableness assistant received higher ratings than the low-agreeableness assistant on six outcomes under both information sources. We detected an HA--LA difference in perceived empathy but no information-source difference in that measure. Perceived intelligence varied with information source and context, but we detected no personality difference; environmental explanations received higher ratings in the emergency scenarios studied here. Extraversion did not produce a source-averaged difference in evaluations, but it moderated how evaluations changed with information sources. These results suggest that effective conversational AI explanations depend on coordinating why, when, and how: selecting an information source appropriate to the context in which the explanation is delivered, while accounting for how that evidence is received under different conversational styles.

\bibliographystyle{ACM-Reference-Format}
\bibliography{cites}

\appendix
\section{Appendix}

\begin{figure*}[ht]
    \centering
    \includegraphics[scale=.43]{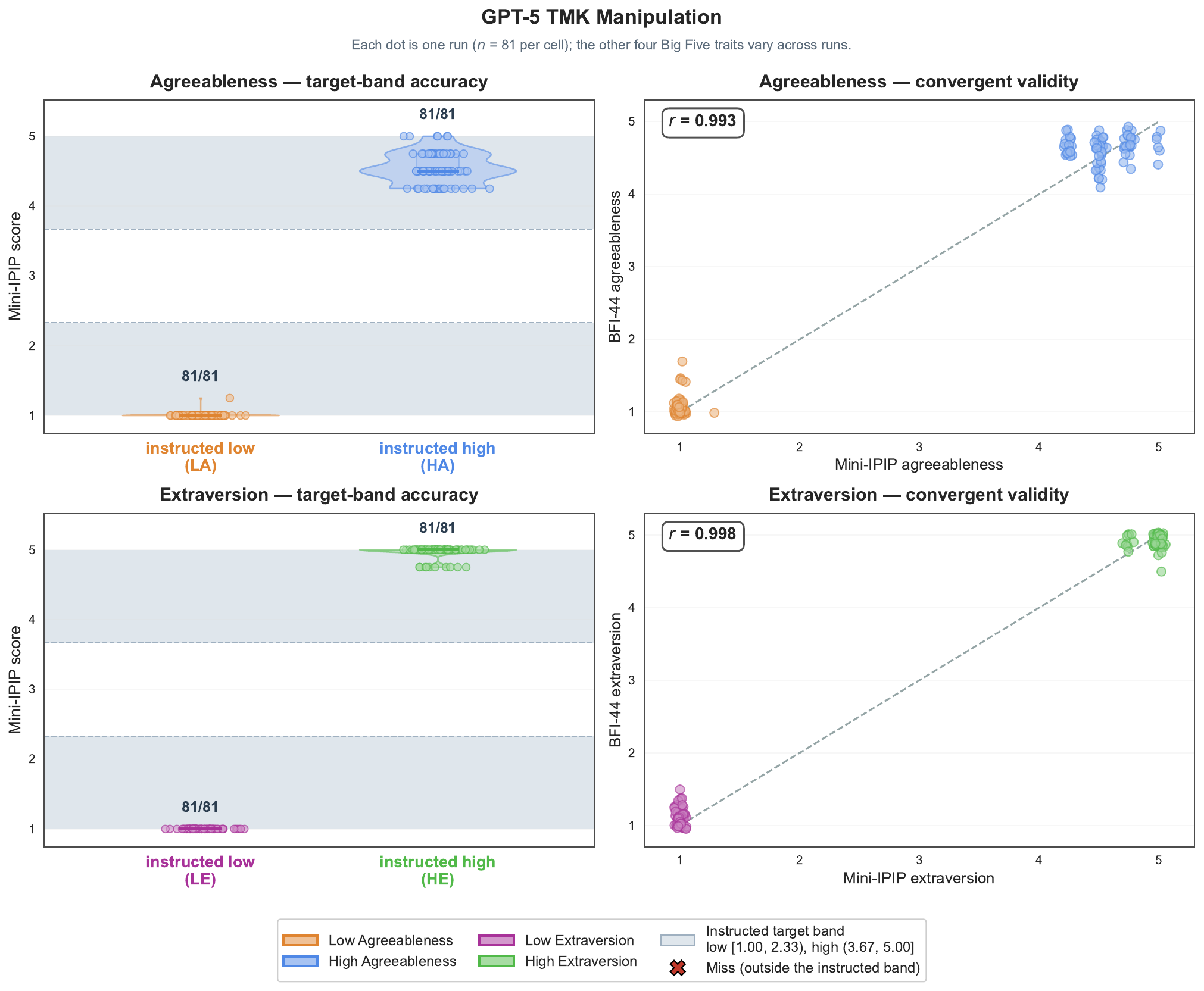}
    \caption{Validation of the Trait Modulation Keys (TMK) used to configure Robin's conversational personality, restricted to the two traits we manipulated. Each dot is one TMK configuration generated with GPT-5 and scored on the Mini-IPIP. Left: Mini-IPIP score for every run against the instructed target band (shaded; low: $[1.00, 2.33)$; high: $(3.67, 5.00]$), with the number of runs inside the band above each violin; the instructed level was recovered in 81 of 81 runs in all four cells. Right: convergent validity between the Mini-IPIP and the BFI-44 for the same runs ($r = .993$ for agreeableness; $r = .998$ for extraversion).}
    \label{fig:tmk}
    \Description{Four panels form a two-by-two grid, with agreeableness on the top row and extraversion on the bottom. The left panels compare instructed low and high trait levels against Mini-IPIP personality scores on a 1–5 vertical axis. Each dot represents one configuration; the other four Big Five traits vary across configurations. Shaded target bands extend from 1 up to, but excluding, 2.33 for low scores, and from above 3.67 through 5 for high scores. Low-condition scores cluster near 1; high agreeableness scores cluster around 4.3–5 and high extraversion scores near 5. Each group is labeled “81/81,” indicating that every configuration falls inside its intended band. The right panels plot Mini-IPIP scores horizontally against BFI-44 scores vertically. Low and high groups form separate clusters near the lower-left and upper-right corners, close to a diagonal reference line. The reported correlations between the two personality measures are .993 for agreeableness and .998 for extraversion.}
\end{figure*}

Table \ref{appendix:desc_personality} reports medians and IQRs for the RQ1 conversational-layer contrasts. Tables \ref{appendix:art_anova_a} and \ref{appendix:art_anova_b} report the complete ART ANOVA results for the mixed-design model, which provides the RQ1 evidential-layer main effects and the RQ2 interaction terms. Table \ref{appendix:baseline_sensitivity} reports the covariate-adjusted endpoint robustness analysis for the LE--HE information-source pattern, and Table \ref{appendix:dissociation} reports the two cross-measure robustness tests. The supplementary materials contain all ART simple effects and descriptives by context and information source. Figure \ref{fig:tmk} shows the TMK manipulation results.

\begin{table*}[h]
\centering
\caption{Covariate-adjusted endpoint robustness analyses within the low- and high-extraversion conditions. The raw-scale model predicts a participant's ENV rating from their UH rating and an LE (vs.\ HE) indicator using HC3 robust standard errors; the rank analysis applies the same model to ranked variables. Positive $b_{\mathrm{LE}}$ indicates a higher adjusted ENV endpoint for LE. We applied BH correction across the seven outcomes within each analysis.}
\label{appendix:baseline_sensitivity}
\begin{tabular}{@{}lrrr@{}}
\toprule
Measure & $b_{\mathrm{LE}}$ & Raw $p_{\mathrm{BH}}$ & Rank $p_{\mathrm{BH}}$ \\
\midrule
Intention to Adopt       & 0.582 & .008 & $<.001$ \\
Empathy                  & 2.276 & .371 & .217 \\
Trust                    & 0.326 & $<.001$ & $<.001$ \\
Reliance                 & 0.253 & .014 & .011 \\
Explanation Satisfaction & 0.296 & .008 & .007 \\
Likeability              & 0.331 & .014 & .012 \\
Intelligence             & 0.121 & .371 & .126 \\
\bottomrule
\end{tabular}
\end{table*}

\begin{table*}[h]
\centering
\caption{Cross-measure tests of differential sensitivity. Each row tests whether a manipulation's standardized effect differs between empathy and perceived intelligence.}
\label{appendix:dissociation}
\begin{tabular}{@{}lrcrr@{}}
\toprule
Test & $\Delta$ & 95\% CI & $p_{\mathrm{boot}}$ & $p_{\mathrm{rank}}$ \\
\midrule
HA--LA: empathy vs.\ intelligence            & $0.78$  & $[0.23, 1.39]$ & .006 & .029 \\
ENV--UH: intelligence vs.\ empathy           & $0.24$  & $[0.13, 0.35]$ & $<.001$ & $<.001$ \\
\bottomrule
\end{tabular}
\vspace{2mm}

\footnotesize \textit{Note.} Measures were z-scored on participant-level scores. The HA--LA row is the difference between Cohen's $d$ values. The ENV--UH row is the difference between standardized within-participant change scores, using the between-person SD as the standardizer. $p_{\mathrm{boot}}$ is the two-sided bootstrap $p$-value from 10{,}000 participant-level resamples; intervals are percentile CIs. $p_{\mathrm{rank}}$ is the Mann--Whitney test for the between-participant contrast and the Wilcoxon test for the within-participant contrast. The rank tests are robustness checks for the two cross-measure comparisons, not additional decision rules.

\end{table*}

\begin{table*}[h]
\centering
\caption{Descriptive statistics across LLM-VA personality conditions (Median and IQR).}
\label{appendix:desc_personality}
\begin{tabular}{@{}lrrrrrrrr@{}}
\toprule
& \multicolumn{4}{c}{Agreeableness} & \multicolumn{4}{c}{Extraversion} \\
\cmidrule(lr){2-5}\cmidrule(lr){6-9}
& \multicolumn{2}{c}{LA} & \multicolumn{2}{c}{HA} & \multicolumn{2}{c}{LE} & \multicolumn{2}{c}{HE} \\
\cmidrule(lr){2-3}\cmidrule(lr){4-5}\cmidrule(lr){6-7}\cmidrule(lr){8-9}
Variable & Median & IQR & Median & IQR & Median & IQR & Median & IQR \\
\midrule
Intention to Adopt & 5.25 & 2.88 & 5.75 & 1.25 & 5.75 & 1.75 & 5.75 & 2.19 \\
Empathy            & 38.16 & 46.27 & 59.32 & 31.06 & 31.85 & 47.99 & 51.80 & 57.34 \\
Trust              & 3.35 & 0.75 & 3.75 & 0.67 & 3.60 & 0.80 & 3.60 & 1.03 \\
Reliance           & 3.45 & 1.20 & 3.85 & 0.70 & 3.60 & 1.22 & 3.80 & 1.08 \\
Explanation Satisfaction       & 3.75 & 0.95 & 4.07 & 0.70 & 4.04 & 0.75 & 3.93 & 0.57 \\
Likeability        & 3.40 & 1.65 & 4.25 & 1.45 & 4.10 & 1.15 & 4.20 & 1.57 \\
Intelligence       & 4.35 & 1.05 & 4.50 & 1.28 & 4.65 & 1.10 & 4.20 & 1.40 \\
\bottomrule
\end{tabular}
\vspace{1mm}

\footnotesize \textit{Note.} Participant-level scores averaged across each participant's four repeated measurements ($n = 35$ per condition). We report medians and IQRs because the analyses are rank-based. Empathy uses a 101-point slider; intention to adopt a 7-point Likert scale; the remaining five measures 5-point scales.
\end{table*}

\begin{table*}[t]
\centering
\caption{ART (aligned-rank transform) ANOVA results for the mixed design (personality condition $\times$ context $\times$ information source; Measures 1--4).}
\label{appendix:art_anova_a}
\begin{tabular}{@{}llrccrr@{}}
\toprule
Measure & Effect & $F$ & $p$ & $p_{\mathrm{BH}}$ & $\eta^2_p$ & Sig. \\
\midrule
\multirow{7}{*}{Intention to Adopt} & Personality & $F(3,136)=2.45$ & .066 & .154 & .051 &  \\
 & Context & $F(1,408)=1.48$ & .225 & .315 & .004 &  \\
 & Source & $F(1,408)=5.25$ & .022 & .031 & .013 & * \\
 & Personality $\times$ Context & $F(3,408)=3.03$ & .029 & .204 & .022 &  \\
 & Personality $\times$ Source & $F(3,408)=3.66$ & .013 & .022 & .026 & * \\
 & Context $\times$ Source & $F(1,408)=11.26$ & $<.001$ & .002 & .027 & ** \\
 & Personality $\times$ Context $\times$ Source & $F(3,408)=0.92$ & .433 & .433 & .007 &  \\
\midrule
\addlinespace
\multirow{7}{*}{Empathy} & Personality & $F(3,136)=3.45$ & .019 & .065 & .071 &  \\
 & Context & $F(1,408)=0.14$ & .708 & .708 & .000 &  \\
 & Source & $F(1,408)=2.40$ & .122 & .122 & .006 &  \\
 & Personality $\times$ Context & $F(3,408)=2.31$ & .076 & .267 & .017 &  \\
 & Personality $\times$ Source & $F(3,408)=1.16$ & .325 & .349 & .008 &  \\
 & Context $\times$ Source & $F(1,408)=2.90$ & .089 & .089 & .007 &  \\
 & Personality $\times$ Context $\times$ Source & $F(3,408)=1.94$ & .123 & .192 & .014 &  \\
\midrule
\addlinespace
\multirow{7}{*}{Trust} & Personality & $F(3,136)=1.78$ & .154 & .202 & .038 &  \\
 & Context & $F(1,408)=2.44$ & .119 & .208 & .006 &  \\
 & Source & $F(1,408)=7.63$ & .006 & .014 & .018 & * \\
 & Personality $\times$ Context & $F(3,408)=1.74$ & .159 & .278 & .013 &  \\
 & Personality $\times$ Source & $F(3,408)=6.42$ & $<.001$ & .002 & .045 & ** \\
 & Context $\times$ Source & $F(1,408)=10.73$ & .001 & .002 & .026 & ** \\
 & Personality $\times$ Context $\times$ Source & $F(3,408)=3.62$ & .013 & .046 & .026 & * \\
\midrule
\addlinespace
\multirow{7}{*}{Reliance} & Personality & $F(3,136)=1.68$ & .173 & .202 & .036 &  \\
 & Context & $F(1,408)=0.38$ & .535 & .625 & .001 &  \\
 & Source & $F(1,408)=17.95$ & $<.001$ & $<.001$ & .042 & *** \\
 & Personality $\times$ Context & $F(3,408)=0.34$ & .798 & .798 & .002 &  \\
 & Personality $\times$ Source & $F(3,408)=5.32$ & .001 & .005 & .038 & ** \\
 & Context $\times$ Source & $F(1,408)=13.15$ & $<.001$ & .001 & .031 & ** \\
 & Personality $\times$ Context $\times$ Source & $F(3,408)=2.05$ & .107 & .192 & .015 &  \\
\bottomrule
\end{tabular}
\vspace{1mm}

\footnotesize \textit{Note.} $\eta^2_p$ denotes partial eta squared. Source = information source. $p_{\mathrm{BH}}$ = Benjamini--Hochberg adjusted across the seven outcome measures within each effect term. Significance codes refer to $p_{\mathrm{BH}}$: $^{***}p_{\mathrm{BH}}<.001$, $^{**}p_{\mathrm{BH}}<.01$, $^{*}p_{\mathrm{BH}}<.05$.
\end{table*}

\begin{table*}[t]
\centering
\caption{ART (aligned-rank transform) ANOVA results for the mixed design (personality condition $\times$ context $\times$ information source; Measures 5--7).}
\label{appendix:art_anova_b}
\begin{tabular}{@{}llrccrr@{}}
\toprule
Measure & Effect & $F$ & $p$ & $p_{\mathrm{BH}}$ & $\eta^2_p$ & Sig. \\
\midrule
\multirow{7}{*}{Explanation Satisfaction} & Personality & $F(3,136)=2.23$ & .088 & .154 & .047 &  \\
 & Context & $F(1,408)=3.46$ & .064 & .149 & .008 &  \\
 & Source & $F(1,408)=18.70$ & $<.001$ & $<.001$ & .044 & *** \\
 & Personality $\times$ Context & $F(3,408)=0.34$ & .794 & .798 & .002 &  \\
 & Personality $\times$ Source & $F(3,408)=3.96$ & .008 & .020 & .028 & * \\
 & Context $\times$ Source & $F(1,408)=16.66$ & $<.001$ & $<.001$ & .039 & *** \\
 & Personality $\times$ Context $\times$ Source & $F(3,408)=1.36$ & .253 & .296 & .010 &  \\
\midrule
\addlinespace
\multirow{7}{*}{Likeability} & Personality & $F(3,136)=4.27$ & .006 & .045 & .086 & * \\
 & Context & $F(1,408)=3.90$ & .049 & .149 & .009 &  \\
 & Source & $F(1,408)=2.47$ & .117 & .122 & .006 &  \\
 & Personality $\times$ Context & $F(3,408)=1.93$ & .124 & .278 & .014 &  \\
 & Personality $\times$ Source & $F(3,408)=1.60$ & .188 & .264 & .012 &  \\
 & Context $\times$ Source & $F(1,408)=11.58$ & $<.001$ & .002 & .028 & ** \\
 & Personality $\times$ Context $\times$ Source & $F(3,408)=4.09$ & .007 & .046 & .029 & * \\
\midrule
\addlinespace
\multirow{7}{*}{Intelligence} & Personality & $F(3,136)=0.11$ & .953 & .953 & .002 &  \\
 & Context & $F(1,408)=6.85$ & .009 & .064 & .017 &  \\
 & Source & $F(1,408)=6.54$ & .011 & .019 & .016 & * \\
 & Personality $\times$ Context & $F(3,408)=1.04$ & .374 & .524 & .008 &  \\
 & Personality $\times$ Source & $F(3,408)=1.10$ & .349 & .349 & .008 &  \\
 & Context $\times$ Source & $F(1,408)=9.59$ & .002 & .002 & .023 & ** \\
 & Personality $\times$ Context $\times$ Source & $F(3,408)=1.85$ & .137 & .192 & .013 &  \\
\bottomrule
\end{tabular}
\vspace{1mm}

\footnotesize \textit{Note.} $\eta^2_p$ denotes partial eta squared. Source = information source. $p_{\mathrm{BH}}$ = Benjamini--Hochberg adjusted across the seven outcome measures within each effect term. Significance codes refer to $p_{\mathrm{BH}}$: $^{***}p_{\mathrm{BH}}<.001$, $^{**}p_{\mathrm{BH}}<.01$, $^{*}p_{\mathrm{BH}}<.05$. RQ1 inference for the conversational layer is based on the two design-defined contrasts on the personality main effect (Table \ref{tab:within_trait_contrasts}); the information-source and context main effects reported here provide the RQ1 evidential-layer inference.
\end{table*}

\end{document}